\documentclass[aps,pre,twocolumn]{revtex4}
\usepackage{amssymb,macros2erev4,graphicx}
\usepackage{times}
\begin{document}
\newcommand{\rme}{{\mathrm{e}}}
\newcommand{\rmd}{\mbox{d}}
\newcommand{\nn}{\nonumber}
\newcommand{\E}{\epsilon}
\newcommand{\half}{\frac{1}{2}}
%

\newif\ifpdf
	\ifx\pdfoutput\undefined
	\pdffalse 
	\newcommand{\fig}[2]{\includegraphics[width=#1]{./figures/#2.eps}}
	\newcommand{\Fig}[1]{\includegraphics[width=\columnwidth]{./figures/#1.eps}}
         \newlength{\bilderlength} 
	\newcommand{\bilderscale}{0.35}
	\newcommand{\storebilderscale}{\bilderscale}
	\newcommand{\bilderskip}{\hspace*{0.8ex}}
	\newcommand{\textdiagram}[1]{%
	\renewcommand{\bilderscale}{0.25}%
	\diagram{#1}\renewcommand{\bilderscale}{\storebilderscale}}
	\newcommand{\diagram}[1]{%
	\settowidth{\bilderlength}{\bilderskip%
	\includegraphics[scale=\bilderscale]{./figures/#1.eps}\bilderskip}%
	\parbox{\bilderlength}{\bilderskip%
	\includegraphics[scale=\bilderscale]{./figures/#1.eps}\bilderskip}}
	\newcommand{\Diagram}[1]{%
	\settowidth{\bilderlength}{%
	\includegraphics[scale=\bilderscale]{./figures/#1.eps}}%
	\parbox{\bilderlength}{%
	\includegraphics[scale=\bilderscale]{./figures/#1.eps}}}
\else
	\pdfoutput=1 
	\newcommand{\fig}[2]{\includegraphics[width=#1]{./figures/#2.pdf}}
	\newcommand{\Fig}[1]{\includegraphics[width=\columnwidth]{./figures/#1.pdf}}
	\newlength{\bilderlength} 
	\newcommand{\bilderscale}{0.35}
	\newcommand{\storebilderscale}{\bilderscale}
	\newcommand{\bilderskip}{\hspace*{0.8ex}}
	\newcommand{\textdiagram}[1]{%
	\renewcommand{\bilderscale}{0.25}%
	\diagram{#1}\renewcommand{\bilderscale}{\storebilderscale}}
	\newcommand{\diagram}[1]{%
	\settowidth{\bilderlength}{\bilderskip%
	\includegraphics[scale=\bilderscale]{./figures/#1.pdf}\bilderskip}%
	\parbox{\bilderlength}{\bilderskip%
	\includegraphics[scale=\bilderscale]{./figures/#1.pdf}\bilderskip}}
	\newcommand{\Diagram}[1]{%
	\settowidth{\bilderlength}{%
	\includegraphics[scale=\bilderscale]{./figures/#1.pdf}}%
	\parbox{\bilderlength}{%
	\includegraphics[scale=\bilderscale]{./figures/#1.pdf}}}
	\pdftrue
\fi

\title{\sffamily\bfseries\large Functional renormalization group
for anisotropic depinning and relation to branching processes}
\author{\sffamily\bfseries\normalsize Pierre Le Doussal{$^1$} and Kay J\"org Wiese{$^2$}\vspace*{2mm}}
\affiliation{{$^1$} CNRS-Laboratoire de Physique Th{\'e}orique de 
l'Ecole Normale Sup{\'e}rieure,
24 rue Lhomond,75231 Cedex 05, Paris France.\\
{$^2$} KITP, University of California at Santa Barbara, Santa Barbara,
CA 93106-4030, USA.\medskip } \date{\small August 9, 2002}

\begin{abstract}
Using the functional renormalization group, we study the depinning of
elastic objects in presence of anisotropy.  We explicitly demonstrate
how the KPZ-term is always generated, even in the limit of vanishing
velocity, except where excluded by symmetry. This mechanism has two
steps: First a non-analytic disorder-distribution is generated under
renormalization beyond the Larkin-length. This non-analyticity then
generates the KPZ-term.  We compute the $\beta $-function to one loop
taking properly into account the non-analyticity. This
gives rise to additional terms, missed in earlier studies.  A crucial
question is whether the non-renormalization of the KPZ-coupling found
at 1-loop order extends beyond the leading one.  Using a
Cole-Hopf-transformed theory we argue that it is indeed uncorrected to
all orders. The resulting flow-equations describe a variety of
physical situations: We study manifolds in periodic disorder, relevant
for charge density waves, as well as in non-periodic disorder. Further
the elasticity of the manifold can either be short-range (SR) or
long-range (LR). A careful analysis of the flow yields several
non-trivial fixed points. All these fixed points are transient since
they possess one unstable direction towards a runaway flow, which leaves open
the question of the upper critical dimension. The
runaway flow is dominated by a Landau-ghost-mode.  For LR elasticity,
relevant for contact line depinning, we show that there are two phases
depending on the strength of the KPZ coupling.  For SR elasticity,
using the Cole-Hopf transformed theory we identify a non-trivial
3-dimensional subspace which is {\em invariant to all orders} and
contains all above fixed points as well as the Landau-mode. It belongs
to a class of theories which describe branching and reaction-diffusion
processes, of which some have been mapped onto directed percolation.
\end{abstract}
\pacs{to be added}
\maketitle

\section{Introduction}\label{intro} The physics of systems driven
through a random environment is by construction irreversible.  The
fluctuation dissipation relation does not hold and one expects the
coarse grained description to exhibit signatures of this
irreversibility. In driven manifolds it has indeed been shown that
non-linear Kardar-Parisi-Zhang (KPZ) terms are generated in the
equation of motion, except when forbidden by symmetry
\cite{Kardar1997,LeDoussalGiamarchi1997}.  A question which was
debated for long time is whether at zero temperature these terms
vanish as the velocity $v \to 0^+$. This is the limit which is relevant
to describe depinning ($f \to f_c^+$). It was found some time ago that
there are two main universality classes for interface depinning
\cite{AmaralBarabasiStanley1994,TangKardarDhar1995,AlbertBarabasiCarleDougherty1998}.
The conclusion was reached mainly on the basis of numerical
simulations, which measure the interface velocity $v(\theta)$ as a
function of an average imposed slope $\theta$, as well as various
arguments related to symmetry.  In the first universality class, the
isotropic depinning class (ID), the coefficient $\lambda$ of the KPZ
term vanishes as $v \to 0^+$ and the KPZ term is thus not needed in
the field theoretic description. In the second class, the anisotropic
depinning class (AD), $v(\theta)$ still depends on $\theta$ as $f \to
f_c^+$ and the KPZ term is present even at $v \to 0^+$.  For AD,
numerical simulations based on cellular automaton models which are believed to
be in the same universality class
\cite{TangLeschhorn1992,BuldyrevBarabasiCasertaHavlinStanleyVicsek1992},
indicate a roughness exponent $\zeta \approx 0.63$ in $d=1$ and $\zeta
\approx 0.48$ in $d=2$.  On a phenomenological level it has been
argued
\cite{TangLeschhorn1992,BuldyrevBarabasiCasertaHavlinStanleyVicsek1992,GlotzerGyureSciortinoConiglioStanley1994}
that configurations at depinning can be mapped onto directed
percolation in $d=1+1$ dimensions, which yields indeed a roughness exponent
$\zeta_{\mathrm{DP}}= \nu_\perp/\nu_{\|} = 0.630 \pm 0.001$,
a dynamical exponent $z=1$, a velocity exponent $\beta_{\mathrm{DP}} =
\nu_{\|} - \nu_\perp \approx 0.636$ and a depinning correlation length
exponent $\nu_{\mathrm{DP}} = \nu_{\|} = 1.733 \pm 0.001$.  Some
higher dimensional extensions of these arguments in terms of blocking
surfaces have been proposed
\cite{StanleyBuldyrevGoldbergerHaussdorfMietusPengSciortinoSimons1992,BuldyrevHavlinStanley1994,HavlinAmaralBuldyrevHarringtonStanley1995,BarabasiGrinsteinMunoz1996}, but there is, to our knowledge, no systematic field theoretical
connection between these problems.

Recently we have reexamined the functional renormalization group (FRG)
approach, introduced previously
\cite{DSFisher1986,NattermanStepanowTangLeschhorn1992,LeschhornNattermannStepanow1996,NarayanDSFisher1992a,NarayanDSFisher1993a}
to describe isotropic depinning to one loop within a $ \epsilon=4-d$
expansion.  We constructed
\cite{ChauveLeDoussalWiese2000a,LeDoussalWieseChauve2002} a consistent
renormalizable field theoretical description up to two loops, taking
into account the main important physical feature --~and difficulty~--
of the problem, namely that the second cumulant $\Delta(u)$ of the
random pinning force becomes non-analytic beyond the Larkin scale.
The 2-loop result for the exponent $\zeta$ shows deviations from the
conjecture\cite{NarayanDSFisher1993a} $\zeta=(4-d)/3$. The reason is
the appearance of ``anomalous'' corrections caused by the non-analytic
renormalized disorder correlator. The 2-loop corrections proved to be
crucial to reconcile theory and numerical
simulations\cite{ChauveLeDoussalWiese2000a,LeDoussalWieseChauve2002}.

The aim of this paper is to extend this FRG analysis to the
universality class of anisotropic depinning. We first show that beyond
the Larkin length, the KPZ-term is indeed generated at $v = 0^+$, as
long as it is not forbidden by symmetry. We explicitly compute the
lowest order corrections for a simple model studied in recent
simulations \cite{RossoKrauth2001b,RossoKrauth2002}. Next we derive
the FRG-flow equations for the second cumulant $\Delta(u)$ in a $4 -
\epsilon$ expansion. In a previous study, Stepanow\cite{Stepanow1995}
considered the model to one loop, but did not take properly into
account the non-analyticity of the renormalized disorder. Since this
is physically important, we 
reexamine the problem
here. Indeed, we find several new important ``anomalous'' corrections,
including the one which generates the KPZ term in the first place,
as well as terms correcting the $\beta $-function. We then introduce an
equivalent description in terms of Cole-Hopf transformed fields. This
description is not only much simpler to study in
perturbation theory (e.g. to two loops it reduces the number of
diagrams by an order of magnitude), but it allows us to obtain a
number of results to {\em all orders}. We argue that the coefficient
$\lambda/c$ which measures the strength of the KPZ non-linearity is
uncorrected to all orders.  We also determine a non-trivial subspace
of the disorder correlators in the form of simple exponentials which
is an exact invariant of the FRG to all orders.  In the Cole-Hopf
variables it is reformulated as the field theory of a specific
branching process, or equivalently reaction-diffusion process.

Our flow-equations allow to study both periodic disorder, relevant for
charge density waves (CDW), and non-periodic disorder, relevant for
lines or interfaces in a random environment. In both cases we find
several non-trivial fixed points. All these fixed points possess at
least one unstable direction and should thus be associated to
transitions. It seems that perturbatively the large scale behaviour is
dominated by a runaway-flow, as it is in the standard KPZ
problem\cite{Laessig1995,Wiese1998a}. The difference is that its
direction is a non-trivial function $\Delta(u)$ in functional
space. Analysis of the above mentioned invariant subspace suggests
that the flow goes towards a specific branching process. The present
RG analysis is however unable to attain the non-perturbative fixed
point. Thus, it also does not allow to strictly decide whether $d=4$
is the upper critical dimension of the anisotropic depinning problem,
which is an open issue.

Finally, since there are indications that KPZ terms may be needed in the
description of the motion of a contact line
\cite{GolestanianRaphael2001}, we have  studied manifolds with
long range elasticity and the simplest KPZ term.  
We determine the critical dimension above
which this KPZ term is irrelevant, as well as the roughness at
crossover. 


\section{Model}\label{Model} We consider a $d$-dimensional interface
(in $d+1$ embedding dimensions) with no overhangs parameterized by a
single component height field $u(x)$.  The case where the disorder is
periodic corresponds to a single component CDW in $d$ dimensions.  The
common starting point is the equation of motion
\begin{equation}\label{start}
\eta \partial_t u_{xt} = c \partial_x^2 u_{xt}  + \lambda (\partial_x u_{xt})^2 + F(x,u_{xt} ) + f_{xt}
\end{equation}
with friction $\eta$, a driving force $f_{xt}=f$ and in the case of
long-range elasticity we replace (in Fourier) $q^2 u_q$ by
$|q|^{\alpha} u_q$ (with mostly $\alpha =1$)
in the elastic force. The pinning force $F(x,u)$ is  chosen
Gaussian with  second cumulant
\begin{equation} \label{2cumul}
\overline{F(x,u) F(x',u')} = \Delta(u-u') \delta^d(x-x')\ .
\end{equation}
Temperature can be taken into account as an additional white noise
$\eta(x,t)$ on the r.h.s.\ of (\ref{start}) with $\langle \eta(x,t)
\eta(x',t')\rangle = 2 \eta T \delta(t-t') \delta(x-x')$, but we will
focus here on $T=0$.

Disorder averaged correlation functions $\overline{\langle
A[u_{xt}]\rangle}= \langle A[u_{xt}]\rangle_S$ and response functions
$\overline{\delta \langle A[u]\rangle/\delta f_{xt}}= \langle
\hat{u}_{xt} A[u] \rangle _{S}$ can be computed from the dynamical action
\begin{eqnarray}
 S&=& \int_{xt} \hat{u}_{xt} (\eta \partial_t 
- c \partial_x^2) u_{xt} - \lambda  \hat{u}_{xt} (\partial_x u_{xt})^2 
\label{msr} \\
&& - \frac{1}{2} \int_{xtt'}
\hat{u}_{xt} \hat{u}_{xt'} \Delta(u_{xt}- u_{xt'}) - \int_{xt} \hat{u}_{xt} f_{xt} \nonumber 
\end{eqnarray}
The uniform driving force $f_{xt}=f>0$ (beyond threshold at $T=0$) may
produce a velocity $v = \overline{\partial_t \langle
u_{xt}\rangle}>0$, a situation which we study by going to the comoving
frame (where $\overline{\langle u_{xt}\rangle}=0$) shifting $u_{xt}
\to u_{xt} + v t$, resulting in $f \to f - \eta v$. This is implied
below: Each $\Delta$ is of the form $\Delta(u_{xt}- u_{xt'} + v(t-t'))$,
 and we always consider the quasi-static limit
$v=0^+$. Perturbation theory is performed both in KPZ and disorder
terms, using the free response function
\begin{equation}\label{R}
\langle \hat u_{q,t'} u_{-q,t} \rangle_0 = R_{q,t-t'}= \eta^{-1}
\rme^{-(t-t') q^2/\eta} \theta(t-t')  .
\end{equation}

\section{Generation of the KPZ-term}\label{genKPZ}
In this section we show how the irreversible (non-potential) KPZ term is
generated, even in the limit $v \to 0^+$, starting from a purely
reversible equation of motion, where all forces are derivatives of a
potential.

Let us first consider the model recently studied numerically by Rosso
and Krauth \cite{RossoKrauth2001b,RossoKrauth2002}, where the elastic
energy is $\int_x E(\nabla u_x)$, and e.g.\ $E(\theta) = \frac{c}{2}
\theta^2 +\frac{c_{4}}{4} \theta^4 $.  The relevant continuum equation
of motion is:
\begin{equation}\label{lf1}
\eta \partial_t u_{xt} = E''(\partial_x u_{xt}) \partial^2_x u_{xt}
+ F(x,u_{xt} + v t) + f - \eta v
\end{equation}
Note first that when $c_4 = 0$, which corresponds to the isotropic
depinning class with $E(\theta) = \frac{c}{2} \theta^2$, the
generation of the KPZ term is forbidden by the statistical tilt
symmetry (STS), i.e.\ the invariance of the equation of motion under a
shift $u_{xt} \to u_{xt} + f_x$ with $f_x = h x$ (or more generally
the covariance under an arbitrary $f_x$) \footnote{If  STS is an {\it
exact} symmetry, e.g. in the continuum limit for a pure $\delta^d(x-x')$
correlator in \ref{2cumul}, or as will be discussed below for models with exact
rotational invariance, the KPZ term is forbidden for any velocity
$v$. If the correlations have a finite range it may not be exact,
leading to a small KPZ term, estimated at finite velocity in
\cite{LeDoussalGiamarchi1997}.}.  When $c_{4} \neq 0$ the model does
not obey STS and the KPZ term is not forbidden, and indeed it is
generated at finite velocity $v>0$. This consideration alone is
insufficient to show that it is still generated as $v \to 0$ since in
that limit the symmetry $u \to - u$ should forbid it. Indeed, if one
performs conventional perturbation theory with an {\it analytic}
disorder correlator $\Delta(u)$, one does immediately find that the KPZ term
vanishes as $v \to 0^{+}$.  However one needs a mechanism by which, as $v \to
0^+$, the symmetry $u \to - u$ remains broken.

As we now show, this mechanism is provided by the non-analytic nature
of the disorder.  We know from studies of isotropic depinning
\cite{NarayanDSFisher1992a,NattermanStepanowTangLeschhorn1992,ChauveLeDoussalWiese2000a,LeDoussalWieseChauve2002} 
that at $T=0$ the coarse grained disorder becomes {\em non-analytic}  (NA)
beyond the Larkin length \footnote{A non-perturbative proof of this
was recently obtained from the exact solution at large $N$
\cite{LeDoussalWiese2001}.}. We show below that this is also the case
for the situation considered here. 

\begin{figure}[t] \centerline{\diagram{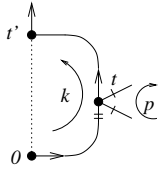}} \caption{{The
diagram generating the irreversible nonlinear KPZ term with one
disorder vertex (notations are as in
Ref.~\cite{ChauveLeDoussalWiese2000a,LeDoussalWieseChauve2002}) and
one $c_4$ vertex (the bars denote spatial derivatives).}}
\label{fig1.a}\end{figure}%
Using the techniques developed in
Ref. \cite{ChauveLeDoussalWiese2000a,LeDoussalWieseChauve2002} the
corresponding perturbation theory, with a non-analytic $\Delta(u)$
becomes (see figure \ref{fig1.a} for notation)
\begin{eqnarray}
\delta \lambda &=&\textdiagram{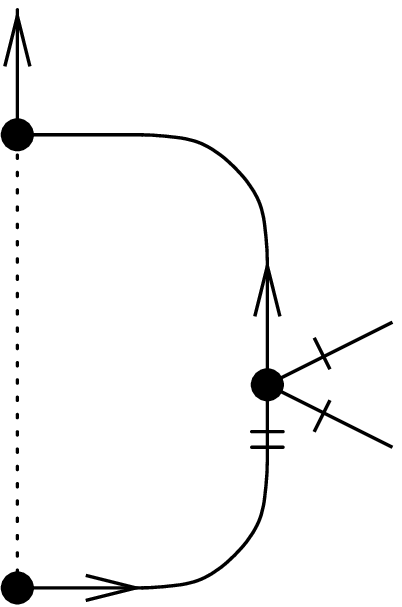}  \nn \\
&=& -\frac{c_{4}}{p^{2}} 
\int_{t>0} \int_{{t'>0}} \int_{k} \rme^{-(t+t')k^2} \left(k^{2}p^{2}+2
(kp)^{2}\right) \nonumber \\
&&\qquad \qquad \qquad \qquad 
\times  \Delta'(u_{x,t+t'}-u_{x,0}+v (t+t')) \ , \nn \\
&&\label{k1}
\end{eqnarray}
At $T=0$, $u_{x,t}$ has vanishing expectation value and the argument
of $\Delta'$ becomes $v (t+t')$. Using that  
\begin{eqnarray}\label{lf1a}
\Delta (u) &=& \Delta (0) + \Delta' (0^{+}) |u| + \half \Delta'' (0^{+})
u^{2}+ \dots\qquad  \\ 
\Delta' (u)&=& \mbox{sign} (u) +  \Delta'' (0^{+}) u+\dots \label{lf2}
\end{eqnarray}
and observing that $t,t'>0$, (\ref{k1}) can be written as 
\begin{eqnarray}\label{lf2a}
\delta\lambda &=& - \frac{c_{4}}{p^{2}} \int_{t}\int_{t'}\int_{k}
\rme^{-(t+t')k^2} 
\left(k^{2}p^{2}+2 
(kp)^{2}\right) \nonumber \\
&&\qquad \qquad  \quad 
\times \left( \Delta' (0^{+}) +  \Delta'' (0^{+}) v (t+t')+O (v^{2}) \right)\nn \\
&&\qquad \label{lf3}
\end{eqnarray}
The leading term of this expansion, which is the only UV-diverging one
for
$4>d>2$, is obtained by setting $v=0$. Integrating over $t,t'$ and
using the radial symmetry in $k$ gives
\begin{equation}\label{lf4}
\delta \lambda = - c_{4} \left(1+\frac{2}{d}
\right)\int_{k}\frac{\Delta' (0^{+})}{k^{2}} + O (v) \ . 
\end{equation}
Similarly, there  is a correction to $c$, which reads
\begin{equation}\label{lf5}
\delta c =c_{4} \diagram{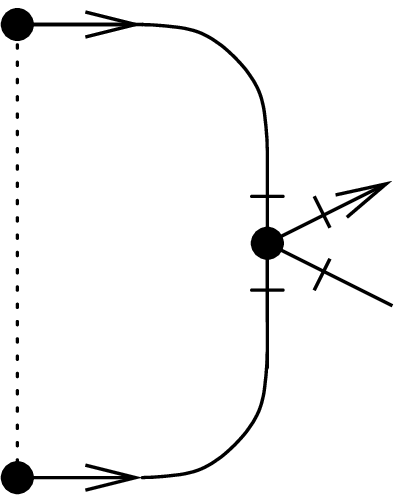} = \int_{k}\frac{\Delta (0)}{k^{2}}
\left(1+\frac{2}{d} \right) 
\end{equation}
leading to 
\begin{equation}\label{lf6}
\delta c = c_{4}\left(1+\frac{2}{d} \right)\Delta (0)
\int_{k}\frac{1}{k^{2}} 
\end{equation}
As will become clear below, the natural coupling for the KPZ-term is
not $\lambda$, but  the ratio $\hat
\lambda = \lambda/c$, which is corrected as \footnote{The equivalence
of the combinatorial factors for $\delta \lambda $ and $\delta c$ can
be deduced by imagining that $\Delta$ be two independent vertices of
the statics and observing that both diagrams correct the same vertex
after using FDT.}:
\begin{equation}\label{lf7}
\delta \hat \lambda = - c_{4} \left(1+\frac{2}{d} \right) (\Delta' (0^{+}) + \hat{\lambda}
\Delta(0) ) \int_{k}\frac{1}{k^{2}} 
\end{equation}
Thus we have shown that the symmetry $u \to - u$ which forbids the KPZ
term (e.g.\ in an analytic perturbation theory where $\Delta'(0)=0$),
is broken here at $v=0^+$ by the non-analytic term, and that a KPZ
term is indeed generated at depinning. As in our previous study
\cite{ChauveLeDoussalWiese2000a,LeDoussalWieseChauve2002} the only
assumption is that the interface always advances forward (or that
backward motion can be neglected in the steady state), supported in
this single component model by no passing theorems
\cite{NarayanDSFisher1992a,RossoKrauth2001b,RossoKrauth2002}.  By
providing a physical mechanism, this explicit calculation confirms the
argument given in \cite{TangKardarDhar1995} based on a Larkin type
estimate of the angle $\theta$ dependence of the  critical force.

Note the sign of the generated KPZ term. Since $\Delta' (0^{+})$ is
negative, $\lambda$ is positive as found in simulations
\cite{AmaralBarabasiStanley1994,TangKardarDhar1995}. It is a bit
counter-intuitive that the surface should become stiffer. Also it
effectively corresponds to the generation of a positive average
curvature. This is presumably through non-analytic coarse grained
configurations of the string (in $d=1 $) since otherwise $\int_0^L
\nabla^2 u = [\nabla u]_0^L$ would grow as $L$ which is unphysical,
while cusps in $u(x)$ allow for such a result.

This model is only a particular case, which shows that the anisotropic
depinning class is rather broad and not limited to anisotropic
disorder. In general, unless they are excluded by symmetry, KPZ-terms
will appear.  One such case, corresponding to a flux line in $1+1$
dimensions which moves perpendicular to itself was considered in
\cite{TangKardarDhar1995}. There disorder is  anisotropic
 with correlators $\Delta_x$ and $\Delta_u$ for
the pinning force. In the case of isotropic disorder
$\Delta_x = \Delta_u$, exact rotational invariance (which in
infinitesimal form reads $u \to u + \theta x$, $x \to x - \theta u$)
should suffice to exclude the KPZ term. We have indeed checked this 
 by adding to the above MSR-action with $\lambda=0$ the non-linear
terms  of 
\cite{TangKardarDhar1995}
\begin{eqnarray} \label{lf8}
 \delta S &=& - \int\limits_{xt} \hat{u}_{xt} \left[A \nabla^2 u_{xt} (\nabla u_{xt})^2 
+ B f (\nabla u_{xt} ^2) \right]  \\
&&\!\!\!\!\!\!\!\!\!\! - \int\limits_{xtt'}\! \hat{u}_{xt} \hat{u}_{xt'} 
\left[  C (\nabla u_{xt})^2 {+} D \nabla u_{xt} \nabla u_{xt'} \right]
\Delta(u_{xt}{-} u_{xt'})  \nonumber \ .
\end{eqnarray}
The generated KPZ term  reads to lowest order
\begin{equation}\label{lf3a}
\delta \lambda=
2 (- A +  C + D) \Delta'(0^+) \int_k \frac{1}{k^2}\ .
\end{equation}
Since the equation of motion of Ref.~\cite{TangKardarDhar1995} for
$\Delta_x = \Delta_h$ corresponds to $A=-1$, $D=C=1/2$, one checks to
lowest order that the KPZ term is indeed not generated. Although
we have not checked it further, it is clear that this property should
extend to all orders. In the
anisotropic class $\lambda$, can a priori be of any sign. The argument
given in \cite{TangKardarDhar1995} suggests that for the flux-line model
$\lambda$ is positive when $\Delta_x < \Delta_h$ and negative for
$\Delta_x > \Delta_h$. Note that anisotropy by itself is not enough  to
 generate the KPZ term, but that a non-linear and non-analytic
disorder correlator is needed, and that this term will of course not be
generated in a simple Larkin-type random force model, where $\Delta_x$
and $\Delta_h$ are  constants.

\section{Dimensional Flory estimates}\label{flory}

Before using analytical methods, let us indicate a simple Flory, or
dimensional, argument which indicates how exponents for ID and AD can
differ. In the absence of a KPZ term and setting $u \sim x^\zeta$ the
two static terms in the equation of motion scale as
\begin{eqnarray}  \label{lf9}
 \nabla_x^2 u  &\sim& x^{\zeta - 2} \\
 F(u_{x} , x) &\sim& x^{- \frac{d+\zeta}{2}}\ . \label{lf10}
\end{eqnarray}
Using $\overline{F(u, x) F(u', x')} \sim \delta(u-u') \delta^d(x-x')$
for random field disorder gives the Imry-Ma value
\begin{equation}\label{lf11}
 \zeta_{\mathrm{F}} = \frac{4-d}{3}
\end{equation}
which can be argued to be exact for the statics and is corrected by
$O(\epsilon^2)$ terms at depinning. These types of arguments typically
give the exact result for LR correlated disorder, as the LR disorder
part is not renormalized. It happens that this range is long enough
for the statics but not for depinning; hence there is a correction at
depinning which increases $\zeta $.  Note that it becomes again exact
for depinning if the range of $\Delta$ in  $u$ or $x$ is large enough
(see e.g. the end of Section IV B in \cite{LeDoussalWieseChauve2002a} and
Appendix \ref{lrdis}).

In presence of a KPZ term the latter scales as 
\begin{equation}\label{lf12}
 (\nabla_x u )^2 \sim x^{2 \zeta - 2}  \ .
\end{equation}
Supposing that it is relevant, it dominates 
 over the elastic term. Balancing KPZ-term against disorder gives the
modified Flory estimate
\begin{eqnarray} \label{FloryKPZ}
\zeta_{\mathrm{F}} = \frac{4-d}{5}\ .
\end{eqnarray}
For $d=1$ it yields
$\zeta_{\mathrm{F}}=0.6$ versus $\zeta=0.63$ observed in
simulations\cite{RossoKrauth2001b}, 
which is not bad an estimate for such a simple argument. Again it is possible that
if one increases the range of $\Delta$ the estimate (\ref{FloryKPZ})
becomes again exact, as is the case for standard KPZ (directed
polymer) see Appendix \ref{lrdis}. Note however that
it works with an upper critical dimension $d=4$, which is an open
question, and is thus merely indicative.

\section{Flow-equations in presence of a  KPZ-term}\label{RG-flow}
Let us start by deriving the FRG flow of $\lambda$, $c$, $\eta$ and
$\Delta$ to one loop starting  from (\ref{msr}).
The KPZ and disorder terms are both marginal in $d=4$ and become
relevant below. Simple dimensional arguments show that these are the
only needed counter-terms. We have computed the effective action to
lowest order. The corrections as given by the diagrams on figure
\ref{fig1.b} are  (for details see
Appendix \ref{app-corrections}):
\begin{equation}\label{results}\hspace{-1cm}
\begin{array}{rcl}
\displaystyle \frac{\delta \eta}{\eta} &=& - \left[
a_{0}c^{-3} \lambda \Delta'(0^+)+c^{-2}\Delta'' (0^{+}) \right] I \\ 
\rule{0mm}{3.8ex}\displaystyle \frac{\delta c}{c} &=& - \left[  a_1
\lambda c^{-3} \Delta'(0^+)  
+ a_2 \lambda^2 c^{-4} \Delta(0) \right] I \\
\rule{0mm}{3.8ex}\displaystyle \frac{\delta \lambda}{\lambda} 
&=& - \left[  a_3 \lambda c^{-3} \Delta'(0^+) + a_4 \lambda^2 c^{-4}
\Delta(0) \right] I  \\
\rule{0mm}{3ex} \displaystyle \delta \Delta &=& \left[  a_5 \lambda^2
c^{-4} \Delta^2  
+c^{-2} ( \Delta'' ( \Delta(0) - \Delta) - (\Delta')^2)\right] I\hspace{-1cm}
\end{array}
\end{equation}
where $I=\int 1/q^4$ (integrated over the shell if using Wilson's
scheme) and the coefficients are:
\begin{eqnarray} \label{lf13}
a_{0}&=&1\ ,\qquad  a_1 = 2(d-2)/d\ , \qquad  a_2 = 4/d \nonumber \\
a_3 &=& a_4 = 4/d\ , \qquad a_5 = 2\ . \label{lf14}
\end{eqnarray}
In the following we will set $d=4$ in these coefficients since they
are universal only to this order. This gives
\begin{equation} \label{lf15}
a_0=a_{1}=a_{2}=a_{3}=a_{4}=1 \ , \qquad a_{5}=2 \ .
\end{equation} One then notes that the quantity $\lambda/c$ remains
uncorrected to first order in $d=4$. In the next section we shall argue
that this remains true to all orders. The corrections to the linear
term in (\ref{msr}) can be interpreted as the correction to the
critical force:
\begin{eqnarray}\label{lf16}
\delta f = - \delta f_c =  (\lambda c^{-2} \Delta (0) + c^{-1} \Delta'
(0^{+})) I_{1} \label{resultsf} 
\end{eqnarray}
where $I_{1}=\int_{q}\frac{1}{q^{2}}$. It does not require an
additional counter-term if we tune $f$ to be exactly at depinning
$f=f_c$.

In view of the non-renormalization of $\lambda /c$ in (\ref{results})
it is useful to denote the unrescaled coupling constants as
\begin{equation}\label{lf17}
 \hat{\lambda} = \frac{\lambda}{c} \ , \qquad 
 \hat \eta = \frac{\eta}{c}  \ , \qquad \hat \Delta = \frac{\Delta}{c^2}\ .
\end{equation}
\begin{figure}[t]
\centerline{\diagram{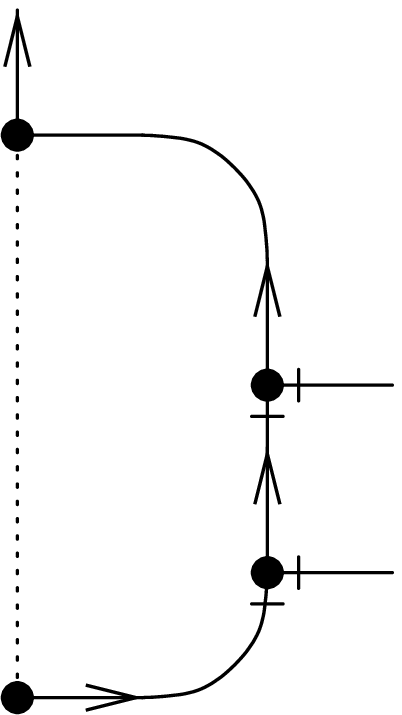}\diagram{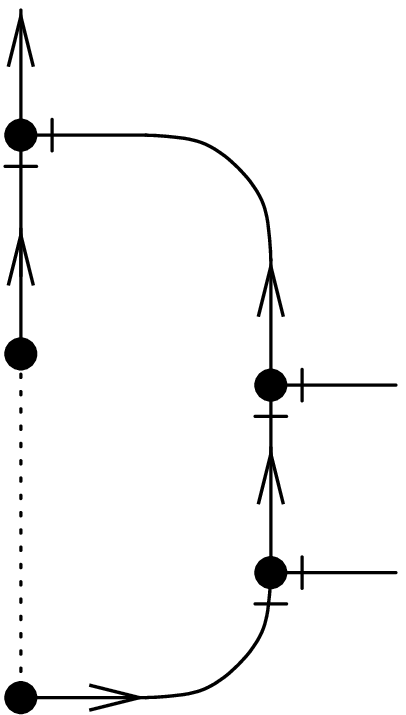}\diagram{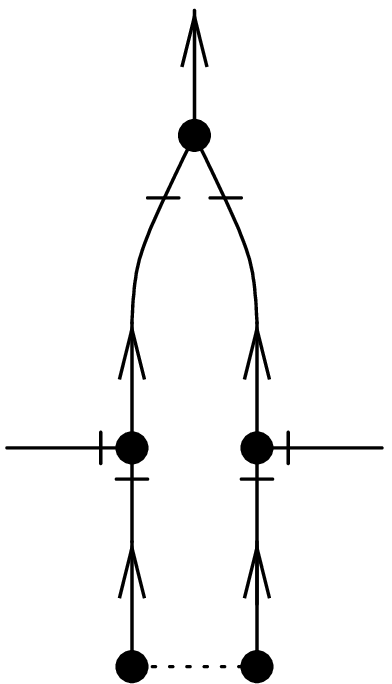}}
\centerline{ \diagram{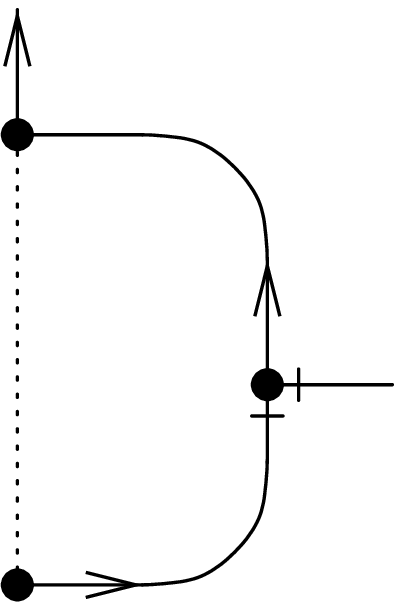}\diagram{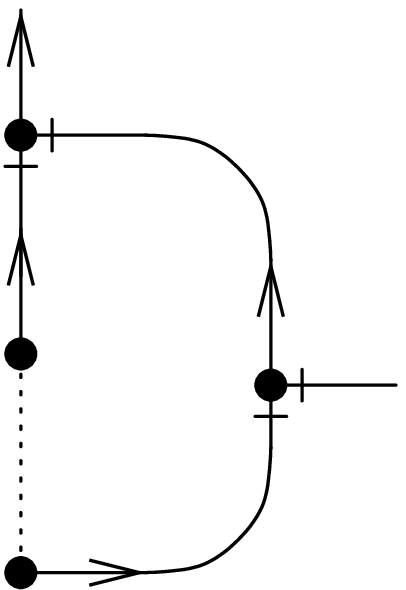}}
\centerline{ \diagram{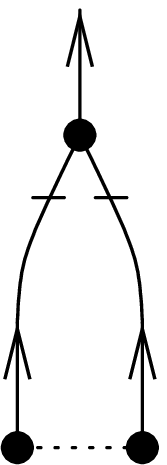}\qquad \diagram{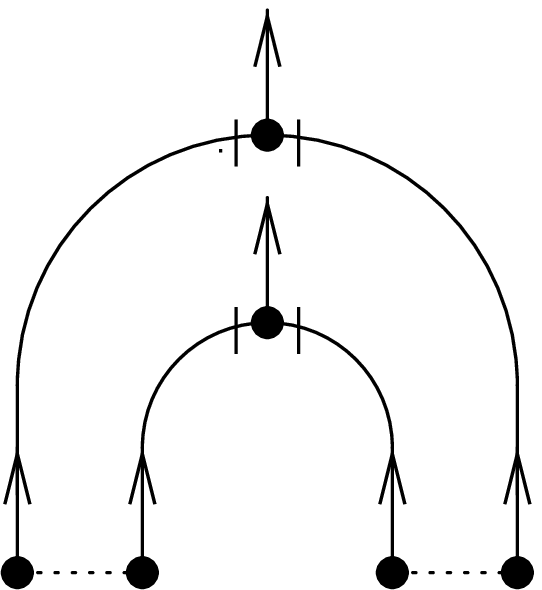} }
\caption{{1-loop-diagrams correcting $\lambda $ (top), $c$
(middle), and $\eta $ and $\Delta (u)$ (bottom).}}
\label{fig1.b}\end{figure}%
One should also notice that if one performs the change of variable in
the initial model $u \to u/\hat \lambda$, $\hat{u} \to \hat{u} \hat
\lambda$, then the free (quadratic) part of the action (proportional to
$c$ and $\eta$) remains invariant while disorder and KPZ terms become:
\begin{eqnarray}
\hat \lambda& \to& 1\nonumber \label{lambda to 1} \\
\Delta(u)& \to& \hat \lambda^2 \Delta(u/\hat \lambda) \label{lf18}
\end{eqnarray}
Thus the coefficient $\hat \lambda$ can be set to one 
upon appropriate redefinitions of  disorder and displacements.

It is natural to start the study of the FRG flow and the search for
fixed points as for $\lambda=0$ by defining the following rescaled
parameters
\begin{eqnarray}\label{lf19a}
 \tilde{\lambda} &=& \hat \lambda \Lambda_\ell^{-\zeta} \\
 \tilde{\Delta}(u) &=& \Lambda_\ell^{2 \zeta - \epsilon} 
\hat \Delta(u \Lambda_\ell^{- \zeta}) \label{lf20}
\end{eqnarray} 
within a Wilson scheme where $\Lambda_\ell = \Lambda \rme^{-\ell}$ is the
running UV cutoff. This yields two coupled equations for the couplings
$\tilde{\lambda}$ and $\tilde{\Delta}(u)$ 
\begin{eqnarray}   \label{lf21}
\partial_\ell \ln \tilde{\lambda} &=&  \zeta  \\
 \partial_\ell \tilde{\Delta}(u) &=&
(\epsilon - 2 \zeta ) \tilde{\Delta}(u) + u \zeta \tilde{\Delta}'(u)
\nonumber \\ 
&& + 2 \tilde{\lambda}^2 \tilde{\Delta}(u)^{2} + [2 \tilde{\lambda}^2
\tilde{\Delta}(0) + 2 \tilde{\lambda }
\tilde{\Delta}' (0^{+})] \tilde\Delta (u)\qquad  \nonumber \\
&& -\tilde{\Delta}' (u)^{2}-\tilde{\Delta}'' (u) (\tilde{\Delta}
(u)-\tilde{\Delta} (0)) \label{beta-1} 
\end{eqnarray}
where here and below we absorb $\epsilon I = S_{4}/(2 \pi)^4$ in the
couplings.  One notes that if there is a fixed point for
$\tilde{\Delta}(u)$, then $\zeta$ is  the roughness
exponent since
\begin{eqnarray} \label{lf22}
\overline{\langle  u_q u_{-q} \rangle } &=& \Delta(0)/c^2 q^4 \nn \\
& =&
\Lambda_\ell^{\epsilon - 2 \zeta } \tilde{\Delta}^*(0)/q^4 \sim
\tilde{\Delta}^*(0)/q^{d-2 \zeta}\qquad 
\end{eqnarray}
when evaluated at scale $\Lambda_\ell=q$.  A more rigorous calculation
uses the effective action\cite{LeDoussalWieseChauve2002} at non-zero
momentum. but to one loop gives the same result.  The dynamical
exponent $z$ in  $t \sim x^z$  and the anomalous dimension of the
elasticity can be determined from
\begin{eqnarray}
- \psi &=& \partial_\ell \ln c = - \tilde{\lambda} \tilde{\Delta}'(0^+)  
-  \tilde{\lambda}^2 \tilde{\Delta}(0)  \label{lf23}\\
z - 2 &=& \partial_\ell \ln (\eta/c) =  -  \tilde{\Delta}''(0^+) 
+ \tilde{\lambda}^2 \tilde{\Delta}(0)\ . \nn \label{lf24}
\end{eqnarray} The correlation-length exponent $\nu $ in 
$\xi \sim (f-f_{c})^{-\nu }$ and
the velocity exponent
$\beta$ in $v \sim (f-f_c)^\beta$
are given by the scaling relations
\begin{eqnarray}
\label{nufromscaling}
\nu &=& \frac{1}{2-\zeta +\psi }\\
\label{betafromscaling}
\beta &=& \nu (z - \zeta) =\frac{z - \zeta}{2 - \zeta + \psi}\ .
\end{eqnarray}
This can be seen by noting that the action (\ref{msr}) is invariant
under $x=\rme^\ell x'$, $t = \rme^{z \ell} t'$, $u = \rme^{\zeta \ell} u'$,
$\hat{u} = \hat{u}' \rme^{(2-z -\zeta -d + \psi) \ell}$ provided $\eta
=\eta'\, \rme^{(2-z + \psi)\ell} $, $c = c'\, \rme^{\psi \ell}$, $\lambda =
\lambda'\, \rme^{(\psi + \zeta)\ell}$, $f = f'\, \rme^{(2 - \zeta + \psi)\ell}$
and $\Delta = \Delta'\, \rme^{(\epsilon - 2 \zeta + 2 \psi)\ell}$ as well
as $T = T'\, \rme^{(2-d-2 \zeta +\psi)\ell}$.  While in presence of STS
one has $\psi=0$, this is not the case here. In a Wilson formulation,
the critical force is obtained by integration over scales of
\begin{equation}\label{lf25}
\partial_\ell f_{c} = - c_{\ell} ( \tilde{\lambda}_\ell
\tilde{\Delta}_\ell(0) + \tilde{\Delta}_\ell' (0^{+}) )
\Lambda_\ell^{2-\zeta} \ ,
\end{equation}
a quantity which physically is likely to remain positive.

A salient feature of the AP class is that the critical force depends
on the angle by which the interface is tilted.  From the arguments of
\cite{AmaralBarabasiStanley1994,TangKardarDhar1995} the characteristic
slope $\theta$ should scale like the ratio of the characteristic
lengths orthogonal and parallel to the interface, $\theta \sim
\xi_\perp/\xi_\parallel \sim (f-f_{c})^{\nu (1-\zeta )}$ 
and more generally the velocity should behave as
\begin{equation}\label{scaling-ansatz}
v(f,\theta) =
(f-f_c(0))^\beta \,
g\!\left(\frac{\theta}{(f-f_c(0))^{\nu(1-\zeta)}}\right) \ . 
\end{equation}
Defining $\lambda _{\mathrm{eff}}$ by\cite{TangKardarDhar1995} $v
(f,\theta )= \lambda _{\mathrm{eff}} \theta^{2}+\dots $, the small
$\theta $ expansion of $v (f,\theta )$ gives he effective
$\lambda_{\mathrm{eff}}$ as
\begin{equation}\label{lf18a}
\lambda_{\mathrm{eff}} \sim (f-f_c(0))^{\beta -2\nu (1-\zeta )} =
(f-f_c(0))^{- \nu (2-\zeta-z)}\ . 
\end{equation}
Performing the redefinition $u=\tilde {u}+\theta x$, we can compute
the critical force as a function of the angle $\theta $ to lowest
order in disorder
\begin{eqnarray}
\delta f_{c}(\theta) &=& - \theta^2 \lambda \left(1 -  \frac{4}{d} I \left( 
\tilde \lambda^2 \tilde \Delta (0) +\tilde \lambda \tilde \Delta'
(0^{+}) \right)\right)\nn\\ 
&=&  - \theta^2 \lambda \left(1 + \frac{\delta \lambda}\lambda \right)
\label{lf26}
\end{eqnarray}
and thus we find an angular dependence, which is increased under
renormalization. 

The notable feature of the above FRG equation is the absence of
corrections to $\hat{\lambda}$ to this order in eqs.\ (\ref{results}).
It is crucial to determine whether this persists beyond one loop. If
there were corrections to higher order this might allow for a
non-trivial fixed point of $\hat{\lambda } $ and thus to fix
$\zeta$. On the other hand, absence of corrections would imply that
for $\zeta>0$, $\tilde{\lambda}$ flows to infinity, which makes the
existence of a perturbative fixed point doubtful. In the next section,
we present a different approach, which
allows to clarify this question.

It is worth noting, that since KPZ-terms are only generated above the
Larkin length, the FRG flow below the Larkin length (as well as the
value of this length) is identical to the case $\lambda=0$. It is
 however instructive to artificially consider the above FRG flow
for an analytic function and with a given imposed bare value of
$\hat{\lambda}$ (setting $\zeta = 0$). One gets
\begin{eqnarray} \label{lf27}
 \partial_{\ell} \tilde{\Delta}(0) &=& \epsilon \tilde{\Delta}(0) + 4
\hat{\lambda}^2 \tilde{\Delta}(0)^2  \\ 
 \partial_{\ell} \Delta''(0) &=& \epsilon \Delta''(0) - 3 \Delta''(0)^2 +
6 \hat{\lambda}^2 \Delta(0) \Delta''(0)\ .\qquad \label{flowdelt2of0}
\end{eqnarray} 
The bare disorder has $\Delta (0)>0$ and $\Delta''(0)<0$. Since all
terms on the r.h.s.\ of (\ref{flowdelt2of0}) have the same sign,
$|\Delta''(0)|$ diverges faster if $\hat{\lambda} \neq 0$, meaning
that the KPZ-term cannot prevent $\Delta (u)$ from becoming
non-analytic. Note that the first equation exhibits a runaway at
$L_{\Delta(0)}$ which can shorten the Larkin length. In $d=4 +
\epsilon$ at $\lambda=0$ there is an unstable fixed point at
$\Delta''(0) = - \epsilon/3$ separating a Gaussian weak-disorder phase
with the bare unrescaled Larkin force producing finite displacements,
and a phase where disorder seems to become non-analytic, only to
become irrelevant at larger scales as can be seen by examining the
flow in the non-analytic space beyond the Larkin length.  At $\lambda
> 0$ there is a fixed line at $\Delta(0) = -\epsilon/(4
\hat{\lambda}^2)>0$ which separates a phase where $\Delta(0)$ grows from
a phase where it decays to zero. On the transition line the flow is
towards  a non-analytic disorder.

\section{Cole-Hopf transformed theory}\label{ColeHopf}
We now introduce the Cole-Hopf transformed theory which has a lot of 
interesting properties. 

Starting from (\ref{start}) we first divide by $c$. This gives
\begin{equation}\label{lf28}
\hat \eta \partial_t u_{xt} =  \partial_x^2 u_{xt}  + \hat \lambda
(\partial_x 
u_{xt})^2 + \frac{1}{c} F(x,u_{xt} ) + \frac{f}{c}
\end{equation}
We then define the Cole-Hopf transformed fields
\begin{equation}\label{lf29}
Z_{xt} := \rme^{\hat \lambda u_{xt}} \qquad \Leftrightarrow  \qquad u_{xt} =
\frac{\ln ( Z_{xt})}{\hat \lambda } \ .
\end{equation}
The equation of motion becomes after multiplying with $\hat \lambda Z_{xt}$
\begin{equation}\label{lf30}
 \hat \eta \partial_t Z_{xt} =  \partial_x^2 Z_{xt}
+ \frac{\hat{\lambda} }{c} F\left(x,\frac{\ln (Z_{xt})}{\hat{\lambda} } \right)  Z_{xt}  + \frac{\hat{\lambda} f}{c}  Z_{xt} 
\end{equation}
and the dynamical action 
\begin{eqnarray}\label{cole}
{\cal S} &=& \int_{xt}\hat{Z}_{xt}\left(\hat \eta
\partial_{t}-\partial _{x}^{2}  \right) Z_{xt} \nonumber \\
&& -\frac{\hat \lambda^{2} }{2 } \int_{xtt'}
\hat{Z}_{xt} {Z}_{xt} \, \hat \Delta\! \left(  \frac{\ln
Z_{xt}-\ln Z_{xt'}}{\hat \lambda }\right)\hat{Z}_{xt'}Z_{xt'} \nn \\
&& - \frac{\hat{\lambda}}{c} f \int_{xt} \hat{Z}_{xt}    Z_{xt}
\end{eqnarray}
It is important to note that the above formal manipulations are only
valid in the mid-point (Stratonovich) discretization. The strategy
therefore is to start from the original equation of motion, which is
interpreted in the It\^o discretization, switch to Stratonovich, make
the change of variables, and then switch back to It\^o.  
Note the identification:
\begin{equation}\label{lf19}
 \hat{u}_{xt} \equiv \frac{\hat{\lambda}}c \hat{Z}_{xt} Z_{xt} 
\end{equation}
and that in this formalism the force (or the distance to the critical force)
corresponds to a mass:
\begin{equation}\label{lf20a}
 m^2 = \frac{\hat{\lambda}}{c} (f -f_c)
\end{equation}
Let us first illustrate how perturbation theory works in this new
formulation and how one can easily recover the 1-loop FRG equation
obtained in the previous section.  Perturbation theory is performed
with the standard response-function. We note a very important
property: To contract $\hat{Z}_{00} $ with a disorder-insertion
$\hat{Z}_{xt} {Z}_{xt} \hat \lambda^2 \, \hat \Delta\!  \left(
\frac{\ln Z_{xt}-\ln Z_{xt'}}{\hat \lambda
}\right)\hat{Z}_{xt'}Z_{xt'}$ and focusing on $Z_{xt}$ (not
$Z_{xt'}$), one can decide to either contract $Z_{xt}$ standing
outside the $\hat \Delta$ or inside. In the first place, this
eliminates the factor $Z_{xt}$, but leaves $\hat \Delta$ underived. In
the second case, deriving the argument of $\hat \Delta$, gives $\hat
\Delta'/\hat \lambda $, together with a factor of $1/Z_{xt}$ from the
inner derivative. The latter also cancels the $Z_{xt} $ standing
outside the $\hat \Delta$. So independently of where one derives, one
always looses the factor of $Z_{xt} $ outside $\hat{\Delta}
$. Contracting $n$ times towards the vertex at $x,t$ thus gives a
factor of $Z_{xt}^{1-n}$. This
observation shows that the diagrammatics are a very simple
generalization of the case without the KPZ-term which was detailed up
to two loops in \cite{LeDoussalWieseChauve2002}. One easily verifies
that the latter case is reproduced upon contracting only the argument
of $\hat \Delta$. To see this, one performs the perturbation theory
and finally takes the limit of $\hat \lambda \to 0$. Each time, one
has contracted a $Z_{xt}$ outside of $\hat \Delta$, one is missing a
factor of $1/\hat \lambda$, and the term vanishes in the limit of
$\hat \lambda \to 0$.  Further remark that for $\hat \lambda \to 0$, the
argument of $\hat \Delta$ becomes
\begin{equation}\label{lf4a}
\frac{Z_{xt}-Z_{xt'}}{\hat \lambda }  = u_{xt}-u_{xt'} + O (\hat \lambda )
\end{equation}
This shows that the perturbation theory for isotropic depinning is
reproduced.

Thus the new diagrams, in the presence of the KPZ-term, can be deduced
from those for $\lambda=0$ by  allowing additional contractions of
a $Z_ {xt}$ outside the $\hat \Delta$. Compared to performing calculations 
using (\ref{msr}) this yields a much simpler perturbation theory,
with far less distinct diagrams. E.g. to two loops, the number of
diagrams is reduced by at least a factor of ten.

Note that now a renormalization of the term $\hat{Z} \hat \Delta Z$ is
allowed, since it is no longer forbidden by STS. Indeed shifting $u_
{xt} \to u_{xt}+\alpha x/\hat \lambda $ and $\hat{Z}_{xt}\to
\hat{Z}_{xt}\rme^{-\alpha x}$, we find that the action changes by
\begin{equation}\label{lf31}
\delta S = \int_{xt} \hat Z_{xt}\left(\alpha^{2}+\alpha \nabla
\right)Z _{xt}\ .
\end{equation}
However, since the action (\ref{cole}) is still translationally
invariant, it remains unchanged under 
\begin{eqnarray}
 Z_{xt} &\to& \mu Z_{xt} \nonumber \\
 \hat{Z}_{xt} &\to& \frac{1}{\mu} \hat{Z}_{xt}\ .\label{lf32}
\end{eqnarray}
Transforming only $ Z_{xt} \to \mu Z_{xt} $ without changing
$\hat{Z}_{xt}$ will allows us later to fix the coefficient of the
Laplacian to unity and transfer all its corrections into corrections
to $\hat{\Delta}$ and $\hat{\eta}$.

We now present the calculations at 1-loop order. 
We start with the corrections to $\hat \eta $. Contracting one disorder
vertex once with itself, we obtain 
\begin{eqnarray}\label{lf5a}
&&\!\!\!\hat \lambda ^{2}  \hat{Z}_{xt'} Z_{xt} \left[\hat \Delta\!\left(\frac{\ln
Z_{xt'}{-}\ln Z_{xt}}{\hat \lambda } \right)+\frac{1}{\hat \lambda }
\hat \Delta'\! \left(\frac{\ln Z_{xt'}{-}\ln Z_{xt}}{\hat \lambda } \right)\!
\right]\nn \\ 
&& \qquad \qquad \times R_{0,t'-t}
\end{eqnarray}
Expanding $\ln Z_{xt'}-\ln Z_{xt}$ for small times yields
\begin{equation}\label{lf6a}
\ln Z_{xt'}-\ln Z_{xt} = \frac{(t'-t)\partial _{t}Z_{xt}}{Z_{xt}} + O
(t-t')^{2} 
\end{equation}
One also has to expand $Z_{xt}$ around $xt'$:
\begin{equation}\label{lf7a}
Z_{xt}= -(t'-t)\partial _{t}Z_{xt'}
\end{equation}
Since the manifold only jumps ahead, the arguments of $\hat \Delta$ and
$\hat \Delta'$ are always positive. Putting all terms together,
we obtain: 
\begin{eqnarray}\label{lf8a}
&&\!\!\! \hat{Z}_{xt'} \partial _{t}Z_{xt'} (t'-t)
R_{0,t'-t} \nn \\
&& \!\times \!\left[\left(\hat \lambda \hat \Delta' (0^{+})+
\hat \Delta''(0^{+}) \right)- \left(\hat \lambda ^{2}\hat \Delta
(0)+\hat\lambda  \hat \Delta' (0^{+})  \right)\right]\qquad \quad 
\end{eqnarray}
Integrating over $t'-t$ yields
\begin{eqnarray}\label{lf9a}
&&\!\!\!\hat{Z}_{xt'} \partial _{t}Z_{xt'} I\nn \\
&&\times 
\left[\left(\hat \lambda \hat \Delta' (0^{+})+
\hat \Delta''(0^{+}) \right)- \left(\hat \lambda ^{2}\hat \Delta
(0)+\hat\lambda  \hat \Delta' (0^{+})  \right)\right]\qquad  \quad 
\end{eqnarray}
We have grouped terms such that in the first bracket there appear the
corrections to $-\frac{\delta \eta }{\eta }$ and in the second those to
$\frac{\delta c}{c}$. Here they appear all together in one diagram. In
the absence of the KPZ-term only the term independent of $\hat \lambda
$ survives. Noting the cancellation between the two terms, we finally
arrive at
\begin{equation}\label{lf33}
\frac{\delta \hat \eta }{\hat \eta } = \left[\hat \Delta''
(0^{+})-\hat \lambda ^{2}\hat \Delta (0) \right] I
\end{equation}

We now turn to corrections to disorder. 
\begin{figure}[b]
\centerline{\Fig{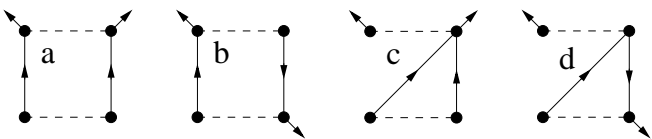}}
\caption{1-loop dynamical diagrams correcting $\hat \Delta$}
\label{1loopDelta}
\end{figure}
\begin{figure}[t]
\centerline{\Fig{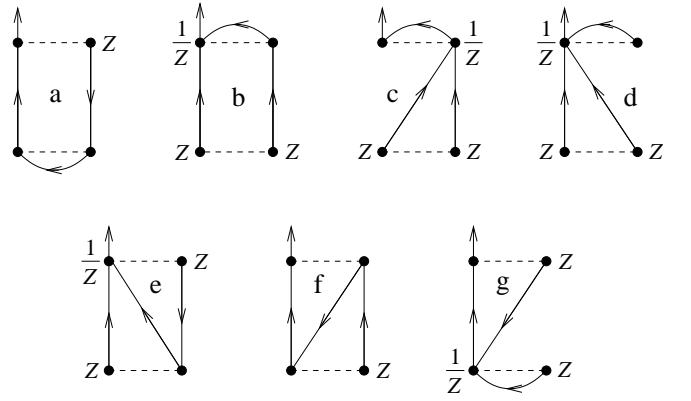}}
\caption{2-loop dynamical diagrams correcting the single $\hat
Z$-component. Diagrams a -- g correct the friction. Only diagrams e
and f have a sufficiently strong divergence in space (after
time-integration) that they can produce spatial gradients. In fact
they both correct $\hat{Z}\Delta Z $. (The diagram is the well-known
sun-set diagram from $\phi ^{4}$-theory.) }
\label{eta-2-loop}
\end{figure}%
Reminding that the arrows can either enter into the argument of
$\Delta$ or into the single $Z$-field, we get the following
contributions (plus some odd terms, which we do not write):
\begin{equation}\label{lf10a}
\begin{array}{rcl}
 \rule{0mm}{3ex} \delta\hat \Delta (u)^{\mathrm{a}} &=& \left[-\hat
\Delta'' (u)\hat 
\Delta (u) + \hat \lambda ^{2}\hat \Delta (u)^{2} \right] I \nn \\
  \rule{0mm}{3ex} \delta \hat\Delta (u)^{\mathrm{b}} &=& \left[- \hat
\Delta' (u)^2  
 + \hat \lambda ^{2}\hat \Delta (u)^{2} \right] I \nn \\
 \rule{0mm}{3ex}  \delta\hat \Delta (u)^{\mathrm{c}} &=& \left[\hat
\Delta'' (u)\hat 
\Delta (0) \right] I \nn \\
  \delta\hat \Delta (u)^{\mathrm{d}} &=& 2 
 \left[ \hat \lambda \hat \Delta (u)  \hat \Delta' (0^{+}) +
\hat \lambda ^{2}\hat \Delta (u)\hat \Delta (0) \right] I
\end{array} 
\end{equation}
These reproduce the corrections obtained in the previous section,
but quite differently.

The Cole-Hopf transformed theory suggests that 
\begin{equation}\label{lf21a}
 \delta \hat{\lambda} = 0
\end{equation}
to all orders. To prove this one has to show that the following terms are
not generated in the effective action
\begin{equation}\label{lf34}
 \hat{Z}_{xt} \frac{1}{Z_{xt}} (\nabla Z_{xt})^2\ .
\label{new}
\end{equation}
It is easy to see that these terms  result from a 
change of $\hat{\lambda}$ (keeping $u_{xt}$ and $\hat{u}_{xt}$ fixed):
\begin{eqnarray}\label{lf35}
Z_{xt} &\to& Z_{xt} \left(1 + \frac{\delta
\hat{\lambda}}{\hat{\lambda}} \ln Z_{xt}\right) \\ 
 \hat Z_{xt} &\to& \hat Z_{xt} \left(1 - \frac{\delta
\hat{\lambda}}{\hat{\lambda}} \ln Z_{xt}\right) \label{lf36} 
\end{eqnarray}
and thus the Laplacian generates (\ref{new}). One can also again
consider a term like $c_{4}$ which is known to produce a shift in
$\hat{\lambda}$ (see (\ref{lf7})), and does produce (\ref{new}) above
together with other irrelevant terms with more gradients. In fact
(\ref{new}) is by power counting the only term marginal in $d=4$ which
can appear.
\begin{figure}[t]
\centerline{\fig{.13\textwidth}{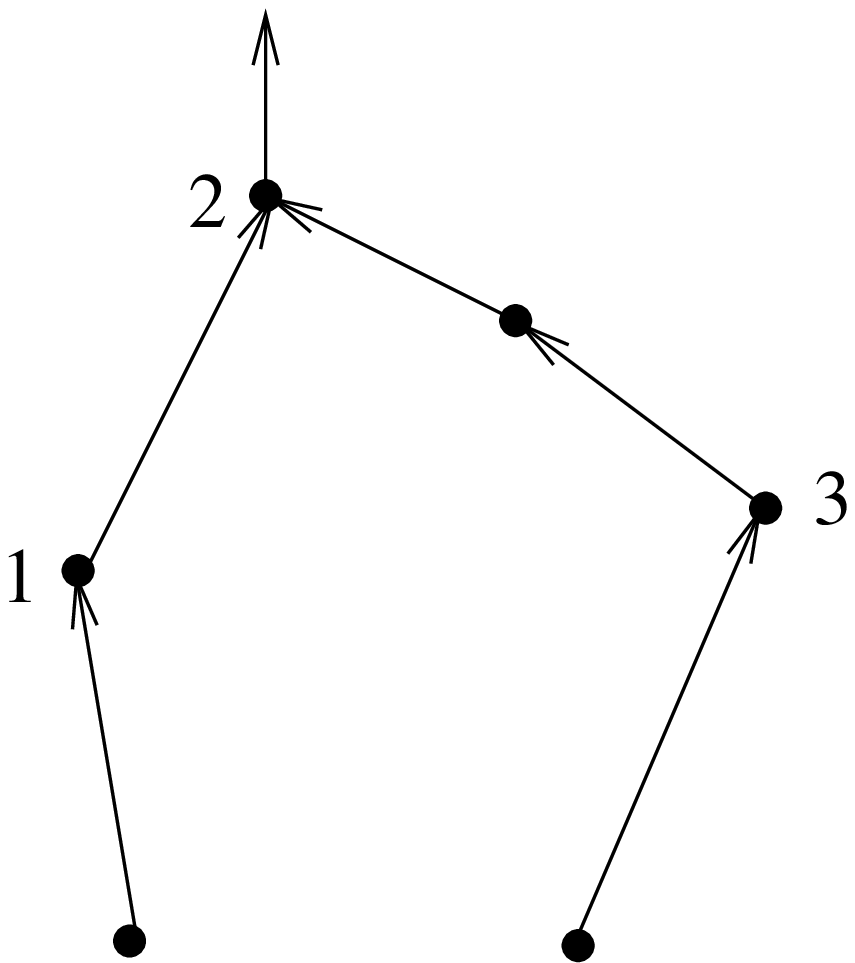}\hfill
\fig{.13\textwidth}{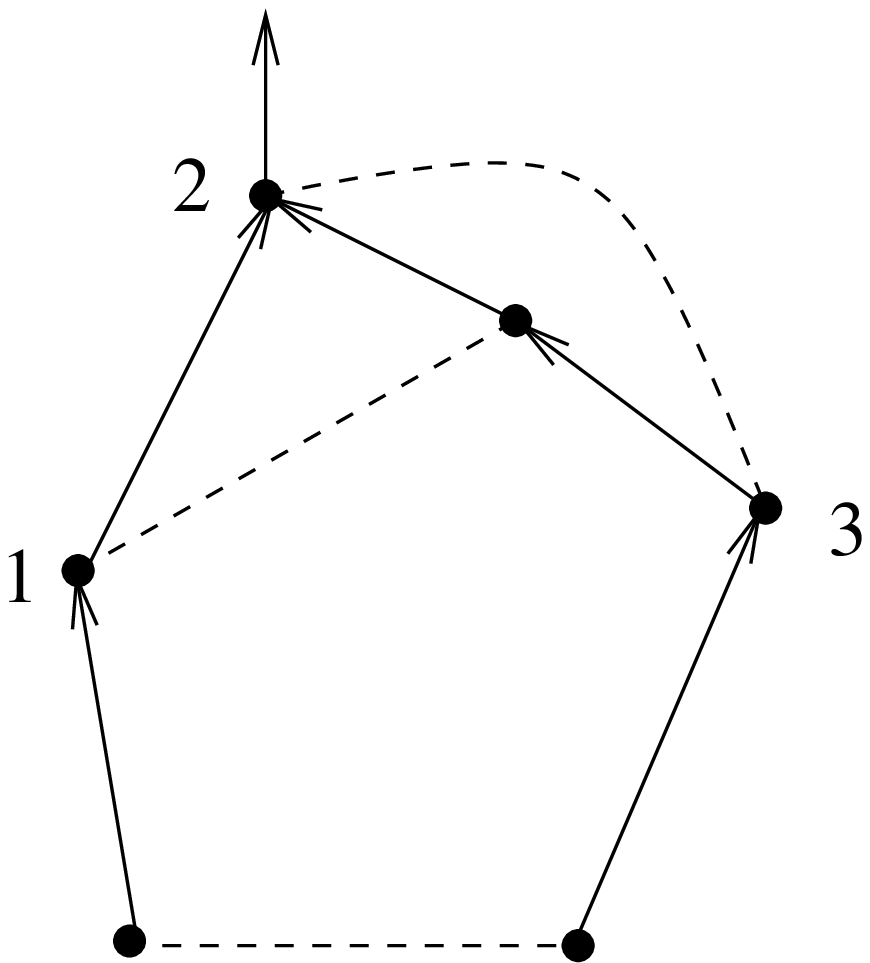}\hfill \fig{.13\textwidth}{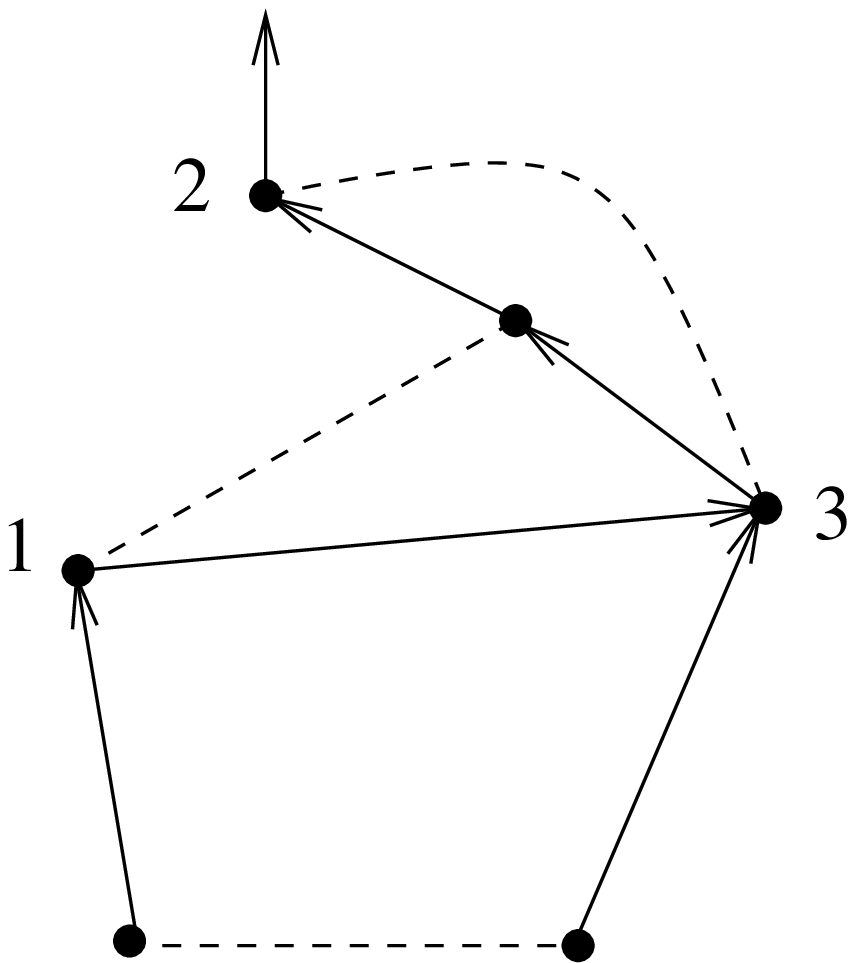}}  
\caption{Figure explaining the non-renormalization of $\hat{\lambda }
$, see main text. }
\label{norenoKPZ}
\end{figure}
This term could in principle come from vertices with several
derivatives acting on $\hat \Delta$ at point $x$. As previously
discussed, it is always compensated, but the compensating factor could
be on a different vertex at position $x'$ and hence produce
(\ref{new}) via a gradient expansion. We have shown in
Fig.~\ref{eta-2-loop} the 2-loop diagrams correcting terms with a
single response field in the effective action and the $Z$ and $1/Z$
fields which appear at each vertex. All terms contribute to $\hat \eta
$.  Graphs $b$, $c$ and $d$ each give a term of the form (\ref{new})
by expanding the $Z^{2}$ on the lower disorder, but the sum of them
cancels. As we will discuss below this is graphically achieved by
moving the ends of the arrows around on the upper vertex, suggesting a
more general cancelation. Another argument is that the divergence in
space between the upper and lower vertex is not strong enough in order
to contribute to (\ref{new}) or $\int \hat{Z} \Delta Z $. For this to
happen, one needs three response-functions between upper and lower
disorder, as is the case for diagrams e and f. They thus both
contribute to $\int \hat{Z} \Delta Z $, but since they have only a
single $Z$ on the lower disorder, they do not contribute to
(\ref{new}).

We now  argue that to all orders in perturbation theory no
diagram proportional to a single $\hat Z$ (one connected component)
can be generated, which contains a factor of $(\nabla
Z)^{2}\frac{1}{Z}$. We believe these arguments to be conclusive;
especially we have not been able to construct any counter-example at
3- or 4-loop order. However the structure of the theory is
sufficiently complicated that some caution is advised.

Look at figure \ref{norenoKPZ}. The response-functions (arrows) in an
arbitrary diagram correcting a single-time vertex have a
tree-structure (left). This diagram can be completed by adding the
disorder-interactions between arbitrary pairs of points (middle). A
potentially dangerous factor of $\frac{1}{Z}$ appears at point
2. Point 2 has a ``brother'' 3, to which it is connected by a disorder
correlator $\hat{\Delta} $ (dashed line). 

Then, two cases have to be distinguished: Either there is no line
entering point 3, then point 3 can contribute his factor of $Z$ to
point 2: Since it is at the same point in space, the difference can be
expanded in a series in time, giving time-derivatives of $Z$ which do
not spoil the argument.

On the other hand, there may be a line entering point 3. This is drawn
on figure \ref{norenoKPZ} (middle). By construction (at least) two
branches (of response-functions) enter at point 2. At least one of
them does not contain the brother of 2 (here point 3). Here it is the
left branch, containing point 1. Now consider the diagram where the
response-function from 1 to 2 is replaced by a response-function from
1 to 3 (right). Since one can always contract last the response-field
at point 1, leading to either the response-function from 1 to 2 or the
one from 1 to 3, these diagrams have the same combinatorial factor,
but differ by a a factor of $-1$, due to the derivative of $\Delta
(\frac{\ln Z-\ln Z'}{\hat \lambda })$ on either the first or the
second argument. This comes in both cases with a factor of
$\frac{1}{Z}$, at the {\em same} position in space but at different
positions in time. However, due to the tree-structure, the
time-integration can always be done freely, and the two vertices
finally cancel. This argument is sufficient before reaching the
Larkin-length. However after reaching the Larking length, the
non-analyticity of the disorder may yield additional sign-functions in
time between both ends of the vertex, as has been observed in
\cite{LeDoussalWieseChauve2002}. Then the proof gets more
involved. There is another very powerfull constraint on the generation
of terms like (\ref{new}): One has to construct a diagram with a
strong spatial ultraviolet divergence, such that after
Taylor-expanding $Z$ in space the additional factor of $x^{2}$
together with this strong ultraviolet divergence gives a pole in $1/\E
$, i.e.\ a logarithmic divergence at $d=4$. This is the situation for
diagrams e and f in figure \ref{eta-2-loop}.  It arises if and only if
there are $2n+1$ response-functions connecting $n$ points in space
(this may well be a sub-diagram), but where response-functions that
connect the same point in space are not counted. In all examples which
we considered up to 4-loop order, which had sufficiently many factors
of $1/Z$, and which had the correct UV-structure, the $(2n+1)$
response-functions where enough to enforce an ordering of times, such
that the mounting proof sketched on figure \ref{norenoKPZ} went
through. We have to leave it as a challenge to the reader to either
find a counter-example or to make the above arguments rigorous.

\medskip

Let us now  return to the analysis of the RG-equations. We
introduce rescaled variables according to 
\begin{eqnarray}\label{lf37}
 \hat \Delta (u) &=& \Lambda_\ell^{ \epsilon - 2 \zeta }
\tilde{\Delta}(u \Lambda_\ell^{\zeta}) \\ 
 \tilde{\lambda}&=& \hat \lambda \Lambda_\ell^{-\zeta }\label{lf38}
\end{eqnarray}
with $\Lambda_\ell = \Lambda  \rme^{-\ell}$. Because we have defined
 $Z=\rme^{\hat \lambda u }$, in order not to generate additional
terms, a rescaling of $u$ demands a (compensating) rescaling
of $\hat \lambda $ such that the product remains unchanged. Even
though this may not be the best choice corresponding to the
existence of a fixed point, it is the only  way to preserve the
Cole-Hopf-transformation, leaving   $Z$ and $\ln Z$  unchanged. The
rescaling of $\hat \Delta$ comes from the rescaling of $\hat \lambda$, which
appears as a factor of $\hat \lambda ^{2} $ in front of $\hat \Delta$ in the
action and as a factor of $1/\hat \lambda $ in the argument of $\hat \Delta$. 

This leads again to the FRG flow equation given in (\ref{beta-1}):
\begin{eqnarray}
\partial _{\ell} \tilde \Delta (u) &=& (\epsilon -2\zeta )\tilde
\Delta (u) + \zeta u 
 \tilde \Delta' (u)\nn \\
&& -\tilde \Delta'' (u)\left(\tilde
\Delta (u)-\tilde \Delta (0) \right) - \tilde \Delta' (u)^2\nn \\
&&+2 \tilde \lambda  \tilde \Delta (u) \tilde \Delta' (0^{+}) \nn \\
&&+2 \tilde \lambda ^{2}\left(\tilde \Delta (u)^{2} +\tilde \Delta
(u)\tilde \Delta (0)\right) \label{beta-2}
\end{eqnarray} 
Further remarkable properties of the Cole-Hopf transformed theory will
be shown below. We now turn to the study of the FRG flow.

\section{Periodic case}\label{PC} We now consider the  case, where
$\hat \Delta (u)$ is a periodic function with period $1$.  The
starting point is (\ref{beta-2}) with $\zeta=0$, thus $\tilde \lambda=
\hat \lambda$ remains constant under renormalization (to all
orders).  Since the period is fixed, $\tilde \lambda$ cannot be scaled
away using (\ref{lambda to 1}). It is thus a continuously varying
parameter and we must study the flow as a function of it.

In eq.
 (\ref{beta-2}) there is a tendency for a runaway flow, as can be
seen by analyzing the flow-equation (\ref{beta-2}) with the trivial
solution $\tilde \Delta(u)=\Delta$
\begin{equation}\label{lf22a}
\partial _{\ell} \Delta = \epsilon  \Delta + 4 \hat \lambda^2 \Delta^2 \ .
\end{equation}
This corresponds to the localization - or self attracting chain - problem studied in
\cite{NattermannRenz1989} and we expect on physical grounds  the
full functional form of  $\Delta(u)$ to be important,
which may lead to other fixed points.

For $\hat \lambda=0$ we already know that there is an unstable fixed
point 
\begin{eqnarray}\label{lf39}
\Delta_{\ell} (u) &=& \Delta^* (u) + c\, \rme^{\epsilon \ell} \\
  \Delta^* (u) &=& \frac{1}{36} - \frac{1}{6} u (1-u)\label{lf40}\ ,
\end{eqnarray}
which describes isotropic depinning for CDW. This fixed point survives
for small $\lambda$ as can be seen from a series expansion in powers
of $\lambda$. Moreover at each order in $\lambda $, $\Delta^{*} (u)$
remains polynomial in $u(1-u)$. We do not reproduce this expansion
here, since we have succeeded in obtaining the fixed point {\em
analytically}. Equation (\ref{beta-2}) possesses the following
remarkable property: 
\begin{itemize}
\item[] {\em A three parameter subspace of exponential
functions forms an exactly invariant subspace.}
\end{itemize}
 Even more strikingly,
this is true {\it to all orders} in perturbation theory. This
property, which is  quite non-trivial, is 
understood in the Cole Hopf theory, as discussed below.

For our purposes, it is more convenient to write
\begin{eqnarray}\label{lf41}
\Delta(u) = \frac{1}{\hat{\lambda}^2} \epsilon f(u \hat \lambda) \ ,
\end{eqnarray}
such that $f$ satisfies the same FRG equation (\ref{beta-2}) with 
$\lambda=\epsilon=1$,  but with period $\lambda$. This allows to
make an ansatz for a family of exponential functions
\begin{equation}\label{f(u)}
f(u) = a + b \, \rme^{- u} + c\, \rme^{u}\ .
\end{equation}
\begin{figure*}[!t] \centerline{\fig{.5\textwidth}{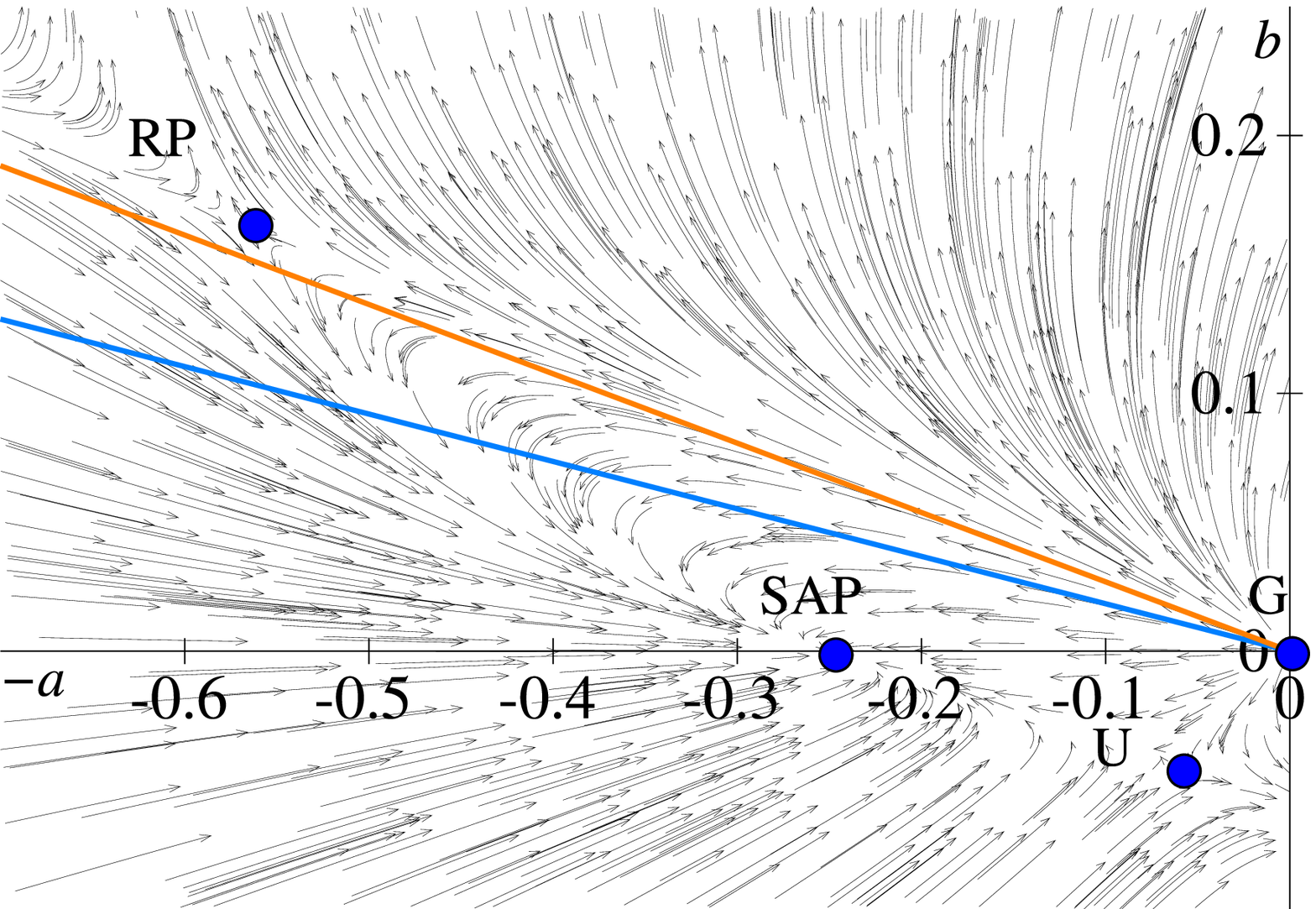}%
\fig{.5\textwidth}{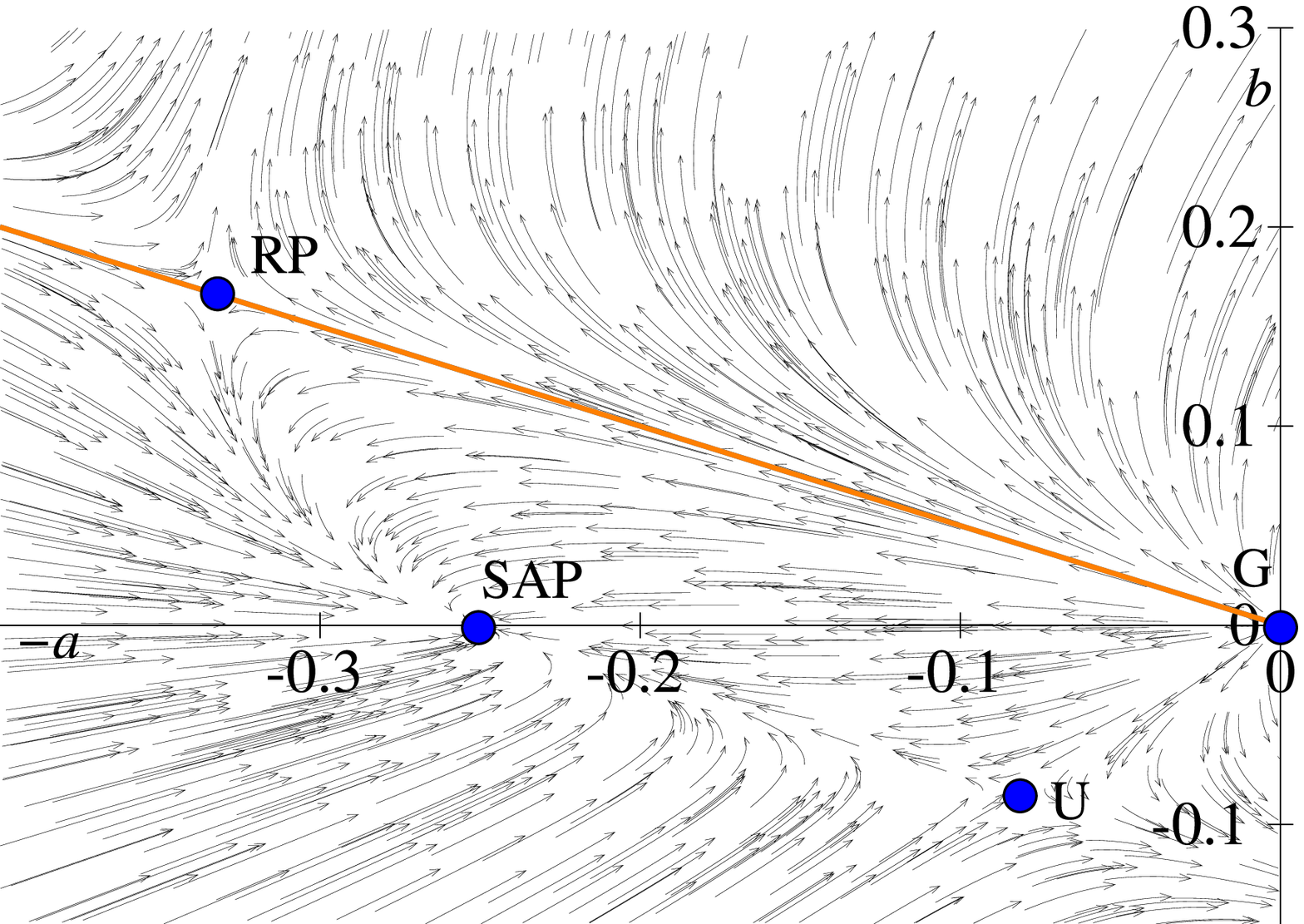}} \vspace*{-.1cm}
\centerline{$\lambda =-1$\hspace*{0.45\textwidth}$\lambda =0$}
\medskip \centerline{\fig{.5\textwidth}{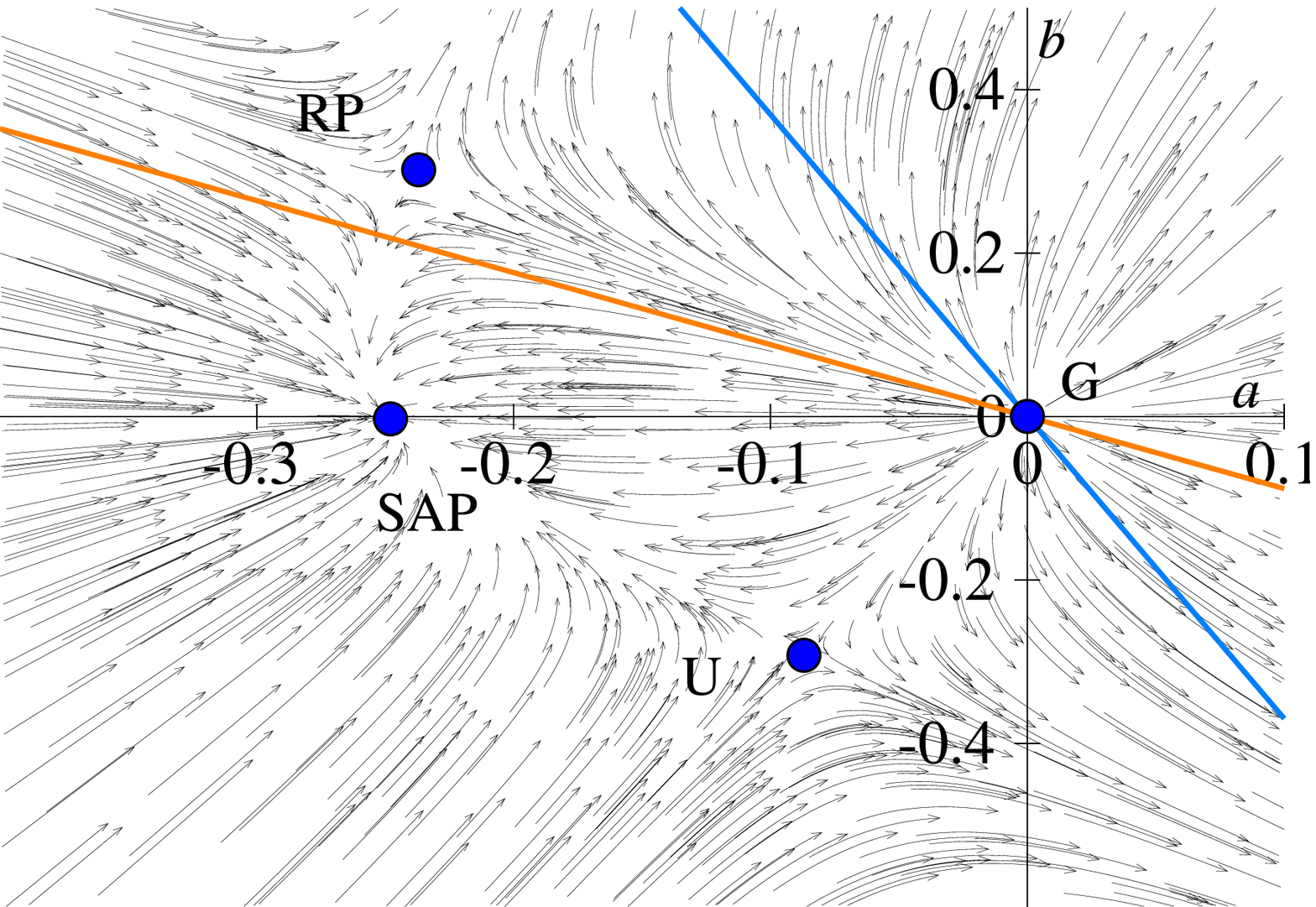}
\fig{.5\textwidth}{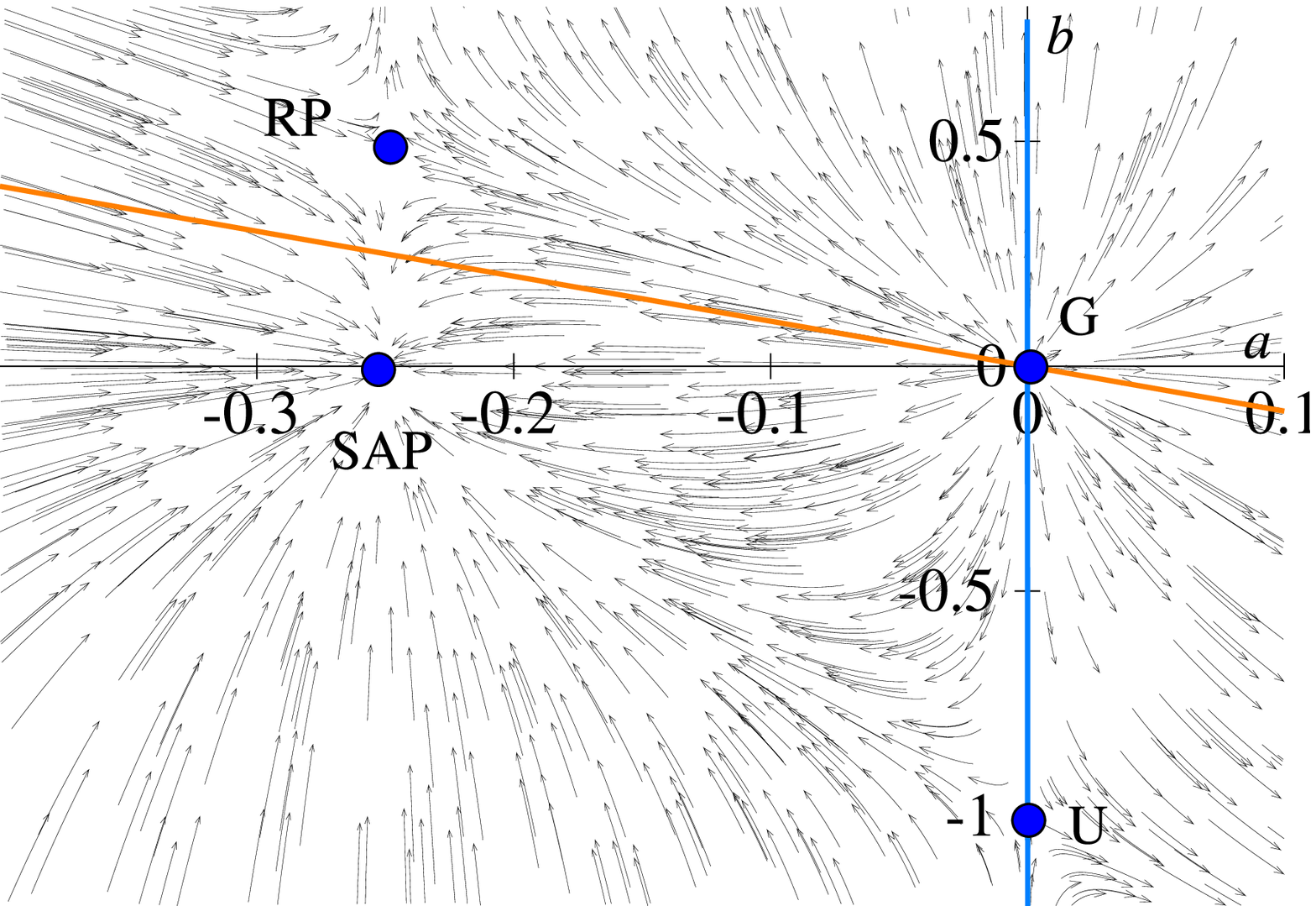}} \vspace*{-.1cm}
\centerline{$\lambda =2$\hspace*{0.45\textwidth}$\lambda =\infty$}
\caption{Fixed point structure for different values of $\lambda$. The
coordinate system is such that $a$ grows to the right and $b$ to the
top. The both separatrices are $b=-\frac{1}{2}a \rme^\lambda$ (blue/dark)
and $b=-a/ (1+ \rme^{-\lambda})$ (orange/bright).} \label{lambda-flow}
\end{figure*}%
The FRG-flow (\ref{beta-2}) closes in this subspace,
leading to the simpler 3-dimensional flow:
\begin{eqnarray}
 \partial_{\ell} a &=& a + 4 a^2 + 4 a c + 4 b c \label{lf42}\\
 \partial_{\ell} b &=& b (1 + 6 a  + b + 5 c ) \label{lf43}\\
 \partial_{\ell} c &=& c (1 + 6 a  + b + 5 c )\label{lf44}
\end{eqnarray}
This works only for amplitude one in the exponential; otherwise higher
modes are generated. Also note that these equations are not symmetric
under the exchange of $b$ and $c$, as one might expect from the
interpretation we will present later.

Requiring periodicity, or equivalently $f(u) = f(\lambda - u)$ imposes
\begin{equation}\label{lf45}
c = b\, \rme^{-\lambda} 
\end{equation}
and one checks that $b/c$ is indeed unrenormalized. Thus one can study
the simpler 2-dimensional flow
\begin{eqnarray}\label{lf46}
\partial_{\ell } a &=& a + 4 a^2 + 4 a b \rme^{-\lambda} + 4 b^2
\rme^{-\lambda} \\ 
\partial_\ell b &=& b (1 + 6 a + b + 5 b \rme^{-\lambda} )
\label{lf47}
\end{eqnarray}
as a function of $\lambda$. A physical requirement is that
\begin{eqnarray}\label{lf48}
 \Delta(0) &=& a + b (1 + \rme^{- \lambda}) > 0 \\
 f_c &\sim& - (\Delta'(0) + \lambda \Delta(0)) \nonumber \\
&=& - \frac{1}{\lambda}
(a + 2 b \rme^{- \lambda}) > 0\ .\qquad\qquad   \label{lf49}
\end{eqnarray}
For $a<0$  this is possible only if  
\begin{equation}\label{lf50}
  - \frac{a}{1 + \rme^{-\lambda}} <  b < - \frac{a}{2} \rme^{\lambda} 
\end{equation}
On the other hand, for  $a>0$ the flow for $a$ is always  $a\to \infty $
in a finite time. Indeed the r.h.s.\ of (\ref{lf46}) is always
positive for $a>0$. For $b>0 $ this is trivial; for $b<0$ this can be
seen from 
\begin{eqnarray}
&&\!\!\!a + 4 a^{2}+ 4 a b \rme ^{-\lambda }+ 4 b^{2 } \rme ^{-\lambda }\nn \\
&& \qquad = 
a + 4 \left(a + b \right)^{ 2} \rme^ {-\lambda }  - 4 a b  \rme^
{-\lambda } + 4 a^{2} (1-\rme^{-\lambda})\nn \\
&&\qquad >0 
\label{lf131}
\end{eqnarray}

The flow given in (\ref{lf46}) and (\ref{lf47}) is shown in figure
\ref{lambda-flow}. There are four fixed points for $d=4 -
\epsilon$. In the original variables they are

\medskip \noindent {\bf (i)} Gaussian fixed point {\tt G} (repulsive in all
directions) with $\Delta(u)=0$.

\medskip \noindent {\bf (ii)} Self-avoiding polymer fixed point {\tt SAP},
where the correlator is a negative constant:
\begin{equation}\label{lf51}
 \Delta(u) = - \frac{\epsilon }{4 \lambda^2} 
\end{equation}
It is the problem of localization in an imaginary random
potential, i.e.\ the Edwards version of the better known self-avoiding polymer. It is
attractive in all directions, even those not drawn here. Writing $f(u)
= -1/4 + \phi(u)$ and linearizing (\ref{beta-2}) gives
\begin{eqnarray}\label{lf52}
 \partial_\ell \phi(0) &=& - \phi(0) - \frac{1}{2} \phi'(0^+)  \\
\partial_\ell \phi' (u) &=& - \frac{1}{2} \phi' (u)\label{lf53}
\end{eqnarray}
This self-avoiding polymer fixed point will not play a role in the
following since for the disordered problem $\Delta(0)>0$. However it
is interesting in other contexts, as discussed below.

\begin{widetext} 

\noindent {\bf (iii)} Fixed point {\tt U}, with one attractive and one 
repulsive direction.
\begin{eqnarray}
 \Delta(u) &=& \frac{1}{\lambda^2} \Bigg[ - \frac{1 + 54
\rme^{-\lambda} + 5 \rme^{-2 \lambda} - (1 + 5 \rme^{-\lambda}) \sqrt{
1 + \rme^{-\lambda} (34 + \rme^{-\lambda} )} }{8
(1 - 5 \rme^{-\lambda} ( \rme^{-\lambda}-8) ))} \nonumber \\
&& \qquad + 
  \frac{2}{ (1 - 7 \rme^{-\lambda} - 3 \sqrt{ 1 + \rme^{-\lambda} (34 
+ \rme^{-\lambda} )}} (\rme^{- \lambda u} + \rme^{- \lambda (1-u)})
\Bigg]\label{lf54} 
\end{eqnarray}
The value at zero 
\begin{eqnarray}
 \Delta(0) &=& - \frac{3 + \rme^{\lambda} (3 + \sqrt{1 + \rme^{-2
\lambda} + 34 \rme^{-\lambda}} )} { 2 \lambda^2 ( 7 + \rme^{\lambda}
(3 \sqrt{1 + \rme^{-2 \lambda} + 34 \rme^{-\lambda}} - 1) ) }
\label{lf55}
\end{eqnarray}
is always negative for $\lambda \geq 0$, thus the FP is unphysical for
our problem in $d=4 - \epsilon$. The combination yielding the
corrections to the critical force 
\begin{equation}\label{lf56}
 f_c = \frac{-1 + \rme^{\lambda} (7 + \sqrt{1 + \rme^{-2 \lambda} +
34 \rme^{-\lambda}} )} { 2 \lambda 
( 7 + \rme^{\lambda} (3 \sqrt{1 + \rme^{-2 \lambda} + 34
\rme^{-\lambda}} - 1) ) }   
\end{equation}
is always positive for $\lambda \geq 0$.

\noindent 
{\bf (iv)} The  random periodic fixed point {\tt RP} has:
\begin{eqnarray}\label{lf57}
\Delta(u) &=& \frac{1}{\lambda^2}
\Bigg[ - \frac{1 + 54 \rme^{-\lambda} + 5 \rme^{-2 \lambda} + 
(1 + 5 \rme^{-\lambda}) \sqrt{ 1 + \rme^{-\lambda} (34 +
\rme^{-\lambda} )} }{8  
(1 - 5 \rme^{-\lambda} ( \rme^{-\lambda}-8) ))} \nonumber \\
&& \qquad + 
  \frac{2}{ (1 - 7 \rme^{-\lambda} + 3 \sqrt{ 1 + \rme^{-\lambda} (34 
+ \rme^{-\lambda} )}} (\rme^{- \lambda u} + \rme^{- \lambda (1-u)}) \Bigg]
\end{eqnarray}
\end{widetext}
\begin{equation}\label{lf11a}
 \Delta(0) = \frac{3 - \rme^{\lambda} (-3 + \sqrt{1 + \rme^{-2
\lambda} + 34 \rme^{-\lambda}} )} { 2 \lambda^2 ( - 7 + \rme^{\lambda}
(1 + 3 \sqrt{1 + \rme^{-2 \lambda} + 34 \rme^{-\lambda}}) ) }
\end{equation}
\begin{eqnarray}
 f_c &\sim& - (\Delta'(0^{+}) + \lambda \Delta(0)) \nn \\
&=& \frac{-7 + \rme^{\lambda} (1 + \sqrt{1 + \rme^{-2 \lambda} + 34
\rme^{-\lambda}} )} { 2 \lambda ( - 7 + \rme^{\lambda} (1 + 3 \sqrt{1
+ \rme^{-2 \lambda} + 34 \rme^{-\lambda}}) ) }\qquad \label{lf58}
\end{eqnarray}
Both quantities $\Delta (0)$ and $f_{c}$ are positive for all $\lambda
\geq 0$, thus this fixed point is physical.

The fixed point {\tt RP} is the continuation of the fixed point
(\ref{lf40}) at $\lambda=0$: Note that apart from a constant only the
term $u (1-u)$ survives from the exponential functions.  Like the
fixed point at $\lambda=0$, it is attractive in one direction (towards
the fixed point {\tt SAP}) and repulsive in another (towards large
$\Delta(u)$). It is thus a critical fixed point.  One can argue that any
perturbation which leads to {\tt SAP} is unphysical,  since at
some scale $\Delta(0)$ becomes negative.  Since we did not find any
strong reason why the system would be exactly on this critical
surface, it is more likely that this FP represents a critical
regime which lies on the boundary of the physical domain.  It is
however interesting that its analytic form can be obtained. In
particular one can compute correlation functions exactly at {\tt RP}.

An important question is whether there are fixed points outside of
the exponential subspace considered above. Let us give a few general
properties. First the flow equations and fixed point conditions near
$u=0$
\begin{eqnarray}
 \partial _{\ell} \tilde \Delta (0) &=& \epsilon \tilde
\Delta (0) + 4 \hat \lambda ^{2} \tilde \Delta (0)^{2}  - \tilde \Delta' (0)^2 + 2 \hat \lambda  
\tilde \Delta (0) \tilde \Delta' (0^{+})\nn \\
 \partial _{\ell} \tilde \Delta' (0^+) &=& \tilde \Delta' (0^+) (\epsilon + 
2 \hat \lambda \tilde \Delta'(0^+) + 6 \hat{\lambda}^2 \tilde \Delta (0)
- 3 \Delta''(0^+) ) \nonumber \\\label{lf59}
\end{eqnarray}
and the flow equation for $\int\tilde \Delta$
\begin{eqnarray}
 \partial _{\ell} \int_0^1 \rmd u\, \tilde \Delta (u)  &=& 
\left( \epsilon + 2 \hat \lambda \tilde \Delta'(0^+) + 2 \hat{\lambda}^2
\tilde \Delta(0) \right) 
\int_{0}^{1}\rmd u\, \tilde \Delta(u) \nonumber \\
&& + 2 \hat \lambda ^{2} \int_0^1 \rmd u\,\tilde \Delta (u)^{2} \label{lf60}
\end{eqnarray}
shows that starting from $\int \tilde \Delta=0$, a positive value for
$\int \tilde \Delta$ is generated in the early stage of the RG. If
there is a fixed point value for $\int \tilde \Delta$ it must be equal
to 
\begin{equation}\label{lf12a}
\int_{0}^{1} \rmd u\, \tilde \Delta^* (u) = -
\frac{ 2 \hat{\lambda}^2  \int_{0}^{1}\rmd u\, \tilde \Delta^*(u)^{2} }{ 
\epsilon + 2 \hat \lambda {{\tilde \Delta}^*}{}'(0^+) + 2 \hat{\lambda}^2
{\tilde \Delta}^*(0) }\label{lf61} \ .
\end{equation}
For small $\lambda$ at least this appears to be negative and of $O(\lambda^2)$.
From the flow-equation for $\Delta' (u)$
\begin{eqnarray}
&&\!\!\! \partial _{\ell} \tilde \Delta' (u) =  - \tilde \Delta''' (u)
\left(\tilde 
\Delta (u)-\tilde \Delta (0) \right)\nn \\
&& + \tilde \Delta' (u) \Big[\epsilon + 
2 \hat \lambda \tilde \Delta'(0^+) + 2 \hat{\lambda}^2 \tilde \Delta (0)
+ 4 \hat{\lambda}^2 \tilde \Delta(u)- 3 \Delta''(u) \Big]  \nn 
\end{eqnarray}
one sees that the behaviour at large $u/\lambda$ must be
exponential. It seems  that there are no non-exponential fixed
points.

The runaway flow will be discussed in the next section.

\section{Random field disorder} Let us now consider non-periodic
functions. The main problem with the natural rescaling of $u = u'
\rme^{\zeta l}$ as in (\ref{lf38}) is that $\tilde{\lambda}$ grows
exponentially, and no fixed point can be found. Let us therefore study
(\ref{lf42})--(\ref{lf44}) setting the rescaling factor
$\zeta=0$.  Again we consider the  invariant subspace of exponential
functions, parameterized by 
\begin{eqnarray}\label{lf62}
\tilde \Delta(u) &=& \frac{1}{\hat{\lambda}^2} \epsilon f(u \hat \lambda) \\
 f &=& a + b \rme^{- u}\label{lf63}
\end{eqnarray}
for $u >0$. Note that we have put the coefficient $c=0$, since we are
not interested in  solutions  growing exponentially in $u$. The flow is
\begin{eqnarray}
 \partial a &=& a + 4 a^2\label{lf64}  \\
 \partial b &=& b (1 + 6 a  + b )\ . \label{lf65}
\end{eqnarray}
The physical requirements now read
\begin{eqnarray}
 \Delta(0) &\sim& a + b > 0      \label{lf66} \\
 f_c &\sim& - a >0   \ .  \label{lf67}
\end{eqnarray}
So it is natural to look in the regime
\begin{equation}\label{lf68}
b > - a \geq 0
\end{equation}
There is again the fixed point
\begin{equation}\label{lf69}
f(x) = - \frac{1}{4} + \frac{1}{2} \rme^{-x}
\end{equation}
which is the infinite $\hat \lambda$ limit of the fixed point {\tt RP}
of the previous section. Since $f (x)$ does not go to zero at infinity
as is expected for random field disorder, and since it is unstable
along the line $a=-1/4$ it is unlikely to have any physical relevance
for the anisotropic depinning class. The other fixed point is
\begin{equation}\label{lf70}
f(x) = - \rme^{-x}
\end{equation}
which has the wrong sign.
One clearly has runaway-flows within the exponential subspace.

We have examined the flow of the FRG numerically. For all initial
conditions considered, which were not exactly at one of the fixed
points mentioned above, we found the solution to explode at some
finite scale, a phenomenon which is known as the {\em Landau
pole}. One issue is to identify the corresponding direction in
functional space.  This issue is related to fixed points in $d=4 +
\epsilon$ dimensions which we now briefly address.  The diagram for $4 +
\epsilon$  is obtained by changing $\Delta \to  - \Delta$ and
$\partial_{\ell} \to - \partial_{\ell}$. This means to replace $a\to -a$ and
$b\to -b$ on figure \ref{lambda-flow} as well as inverting the
direction of all arrows.   {\tt U} then 
controls the boundary between the strong coupling regime of KPZ and
the Gaussian fixed point {\tt G};  {\tt SAP} 
between localization (attractive polymers); the Gaussian fixed point is
multi-critical and {\tt RP} between branched polymers and Gaussian.
For the random field case we now have 
\begin{equation}\label{lf73}
\Delta(u) = \frac{1}{\hat{\lambda}^2} \epsilon f(u \hat \lambda) 
\end{equation}
The  fixed point {\tt RP} gives
\begin{equation}\label{lf74}
f(x) = \frac{1}{4} - \frac{1}{2} \rme^{-x}
\end{equation}
and the fixed point {\tt U} is
\begin{equation}\label{lf75}
f(x) = \rme^{-x}\ ,
\end{equation}
which has the correct sign. It has a vanishing critical force, but is
a good candidate for the critical behaviour between the Gaussian phase
and the strong coupling KPZ phase.

Let us now study the runaway flow for $d=4 - \epsilon$. Suppose that
$\Delta_{1} (u)$ is the solution of the $(4+\epsilon )$-dimensional
flow equation at $\epsilon =1$. Then  
\begin{equation}
 \Delta_{\ell} (u):= g_{\ell}  \Delta_{1} (u)\label{lf76}
\end{equation}
leads to the flow-equation for the amplitude $g_{\ell}$
\begin{equation}
 \partial g_{\ell} = \epsilon g_{\ell} + g_{\ell}^2\label{lf77}\ .
\end{equation}
For the RF-case one has one such point at the boundary of the physical
domain, as can be seen from the flow-equations 
\begin{eqnarray}
a &=& 0 \label{lf78}\\
 \partial b &=& b + b^2\label{lf79}
\end{eqnarray}
Also not that since this mode explodes after a finite renormalization
time, it is difficult to avoid.  However, we have not yet completely
ruled out another scenario, where at least some trajectories have
exponential growth. Making the ansatz
\begin{equation}\label{lf71}
 \Delta_l(u) = \epsilon ( \rme^{2 \zeta l} f(u) + g(u) )\ ,
\end{equation}
this requires to find a solution to the $\beta $-function at $\epsilon
=0$, which we write symbolically
\begin{equation}\label{lf72}
 \beta(f,f)=0 \ .
\end{equation}
One can check that near zero such a solution is in principle
possible. There is a solution, which vanishes at $u=u^* = 1.39895$
(for $\lambda=1$) and becomes negative beyond. One can argue that one
needs it only up to $u=u_{0}<u^{*}$, since the linear term can no
longer be neglected when $f (u)$ approaches zero. Noting
$r=(1+\sqrt{5})/4$ one has $f'(0)=-1/r$, $f''(0)=(1+ 2 r^2)/(3
r^2)$. In this scenario $\zeta$ is determined together with $g$.  It
is unclear how this carries to higher orders, since it seems to
require that $f (u)$ is also solution of the $\beta $-function at
$\epsilon =0$. This is however exactly what happens in the case
$\lambda=0$ with the constant shift $\Delta(0)$. Although numerics
does not seem to confirm it, it is hard to disprove.  A
question which remains to be answered is what  the basin of
attraction of  runaway growth and eventually of exponential growth are.

\section{General arguments from the Cole-Hopf representation and 
branching processes}
\begin{figure}[b]
\centerline{\Fig{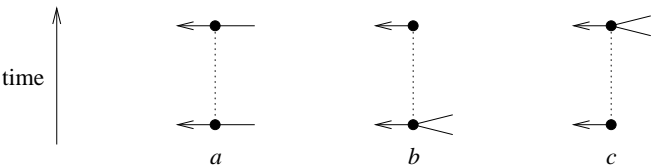}} \caption{The three vertices proportional
to $a$, $b$ and $c$ in 
$\int_{x,t<t'}\hat Z_{xt}\hat Z_{xt'}\left(a
Z_{xt}Z_{xt'}+bZ_{xt}^{2}+cZ_{xt'}^{2} \right)$.}
\label{f:branch}
\end{figure}
In the Cole-Hopf representation, it is easy to see why the
exponential manifold is preserved to all orders. Let us insert
\begin{equation}\label{lf80}
\Delta(u) = \frac{1}{\hat \lambda^2} \left(a + b\, \rme^{- \lambda u} + c\,
\rme^{\lambda u}\right)
\end{equation}
in (\ref{cole}).
The complicated functional disorder takes a very
simple polynomial form  
\begin{eqnarray}\label{c2}
{\cal S} &=& \int_{xt}\hat{Z}_{xt}\left(\hat \eta
\partial_{t}-\partial _{x}^{2}  \right) Z_{xt} \nonumber \\
&& - \int_{x}\int_{t<t'}
\hat{Z}_{xt}\hat{Z}_{xt'} \left(a
Z_{xt}Z_{xt'}+bZ_{xt}^{2}+cZ_{xt'}^{2} \right)\ .\qquad \  
\end{eqnarray}
Note that
we have ordered the vertices in time to distinguish between
$b$ and $c$ taking correctly into account that the
full correlator for the present non-analytic
e.g. random field, problem is (\ref{lf80}) with $u$ replaced by $|u|$
(if (\ref{lf80} holded as an analytic function there would be no
distinction between $b$ and $c$, ``thus no  arrow of time''). 

The vertices presented on figure \ref{f:branch} can be interpreted as
branching processes, and we shall thus call this form {\em branching
representation}. 
\begin{figure}[t]
\centerline{\Fig{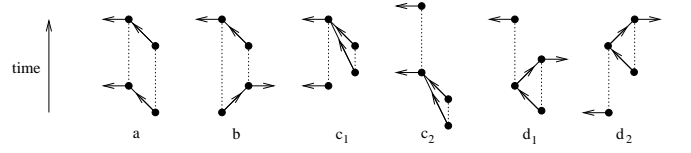}}
\caption{Diagrams correcting the disorder in the
branching-representation. }
\label{f:Dob}
\end{figure}
Let us show how one reproduces the flow-equations (\ref{lf42})-
(\ref{lf44}). In the time-ordered representation, diagrams a to d of
figure \ref{1loopDelta} have the form given on figure \ref{f:Dob}. To
simplify notations, we set $\lambda =1$. Then
\begin{eqnarray}
\Delta (u)&=& a+ b \rme^{-u} + c\rme^u\label{lf132} \\
\Delta' (u)&=&  -b \rme^{-u} + c\rme^u\label{lf133} \\
\Delta''(u) &=&   b \rme^{-u} + c\rme^u\label{lf134} \\
\Delta (0)&=& a+ b  + c\label{lf135}\\
\Delta' (0^{+})&=&c-b\label{lf136}\\
\Delta''  (0^{+})&=& b+c \label{lf137}
\end{eqnarray}
The diagrams have the following contributions
\begin{eqnarray}
\delta \Delta^{\mathrm{a} } (u) &\longrightarrow& \left\{\begin{array}{rcl}
\delta a &=& aa I \\
\delta b &=& ab I \\
\delta c &=& ac I
\end{array}  \right. \label{lf138} \\
\delta \Delta^{\mathrm{b} } (u) &\longrightarrow& \left\{\begin{array}{rcl}
\delta a &=& ( aa + 4 bc) I \\
\delta b &=& 2 b a I \\
\delta c &=& 2 a c I
\end{array}  \right. \label{lf139}\\
\delta \Delta^{\mathrm{c} } (u) &\longrightarrow& \left\{\begin{array}{rcl}
\delta a &=& 0 \\
\delta b &=& 2 b (a+b+c ) \frac{I}{2} \qquad \mbox{(from
$\mathrm{c}_2$) \qquad }\\ 
\delta c &=&  2 c (a+b+c ) \frac{I}{2}\rule{0mm}{2.5ex} \qquad
\mbox{(from $\mathrm{c}_1$)}
\end{array}  \right.\label{lf140} \\
\delta \Delta^{\mathrm{c}} (u) &\longrightarrow& \left\{
\begin{array}{rcll}
\delta a &=& 2 a (a+2b) \qquad& \mbox{(from  $\mathrm{d}_1$ and
$\mathrm{d}_2$)} \\ 
\delta b &=& 2 b (a+b+c ){I} \qquad & \mbox{(from $\mathrm{d}_2$)}\\
\delta c &=&  2 c (a+b+c ) {I} & \mbox{(from $\mathrm{d}_1$)}
\end{array}  
\right. \label{lf141} 
\end{eqnarray}
Note that the factors of $2 $ come in general from contracting
$Z_{xt}^{2}$. The non-trivial factor of $\frac{1}{2}$ is due to the
fact that the two right-most points in $\mathrm{c}_{1}$ and
$\mathrm{c}_{2}$ are time-ordered. To relate the integral to $I$, one
can first symmetrize (yielding the factor of $\frac{1}{2}$) and then
freely integrate over time. Also note that only the last diagram,
$\mathrm{d}_{1}+\mathrm{d}_{2}$ contributes to the asymmetry between
$b$ and $c$. 

In the same way, one can reproduce the corrections to $\eta $.
The only vertex in (\ref{c2})  which contributes at leading order is the one
proportional to $a$: $b$ does not allow for a contraction and $c$ will
have both $\hat{Z} $ and $Z$ at the same point, thus only corrects the
critical force. $a$ leads to 
\begin{equation}\label{lf142}
\hat{Z_{xt}} Z_{xt'} R_{x,t-t'} 
\end{equation}
and after a gradient-expansion following the procedure described after
(\ref{lf5a}) we have
\begin{equation}\label{lf143}
\hat{Z}_{xt} \dot Z_{xt} (t'-t) R_{x,t'-t} \ .
\end{equation}
Integration over $t'$ leads to the correction to $\hat{\eta } $
\begin{equation}\label{lf144}
\frac{\delta \hat{\eta } }{\hat{\eta } } = - a I \ ,
\end{equation}
which is the same one obtained from (\ref{lf33}) using (\ref{lf135})
and (\ref{lf137}). 

Let us now exploit this representation further: It is immediately clear,
that one cannot generate $\rme^{- 2 \lambda u}$ which corresponds to
\begin{equation}\label{lf82}
\int_{x}\int_{t<t'} \tilde{Z}_{x t} \tilde{Z}_{x t'} \frac{Z_{x
t'}^3}{Z_{x t}}
\end{equation}
or any other such fractions. This shows that the space of functions
spanned by (\ref{lf80}) is indeed closed to {\em all orders in
perturbation theory}. Also there is no renormalization to $\hat
\lambda$, whereas a correction to the elasticity $\int \hat{Z} \partial
^{2}_{x}Z $ is allowed, and indeed shows up at 2-loop order. 
\begin{figure}[t]
\centerline{\Fig{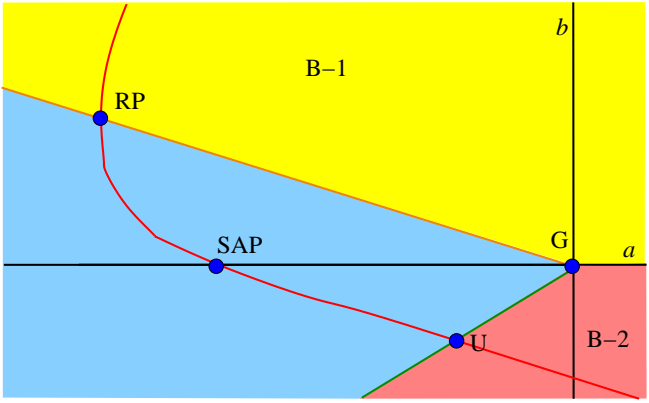}}
\caption{The three phases of the flow-diagrams on figure
\ref{lambda-flow}. }
\label{f:phases}
\end{figure}
\begin{figure}[b]
\centerline{\fig{\columnwidth}{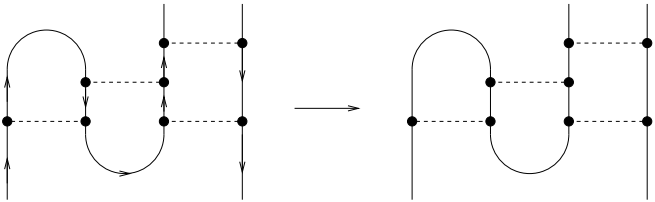}}
\caption{A self-avoiding polymer.}
\label{f:sap}
\end{figure}

Finally, note that the domain of variation of $u$, in the periodic
case yields  an action with multiplicative periodicity in
$Z$, but this does not seem to be important here. 

Let us now discuss the relation of our findings with self-avoiding
polymers, branching
processes and directed percolation.

First, on figure \ref{f:sap} we have drawn a diagram corresponding to
the perturbation expansion of fixpoint {\tt SAP}, which is the only
fully attractive fixed point in the phase-diagram \ref{f:phases}.
One easily checks that by integrating over times, one recovers a
standard $\phi ^{4}$-perturbation theory, as depicted on figure
\ref{f:sap}. By first integrating over the momenta, one recovers the
perturbation expansion of self-avoiding polymers. It is well known,
that this fixed point is stable. In terms of particles, it can be
interpreted as the world-lines of diffusing particles, which are not
allowed to visit twice the same point in space. Let us now add some
terms $b$ and $c$. In interesting limit is $\lambda =\infty $, since
there $c$ can be set to zero. Adding a term propotional to $b$, the
diffusing particle is allowed to branch. More precisely, two particles
can meet at a time $t$. Then one of the particles becomes inactive,
before reappearing at some later time $t'>t$. One can interprete this
as 
\begin{figure}[t]
\centerline{\fig{\columnwidth}{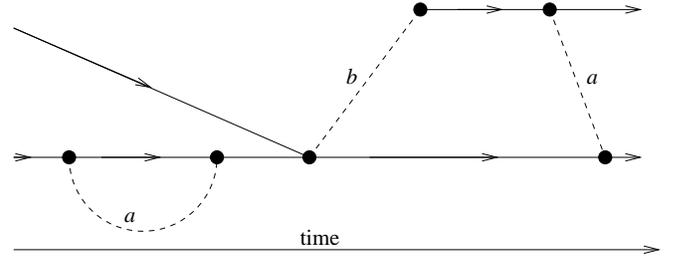}}
\caption{Self-avoidance plus branching.}
\label{f:aab-branch}
\end{figure}
\begin{eqnarray}\label{lf145}
\mathrm{A} + \mathrm{A}~ &\longrightarrow&~ \mathrm{A} + \mathrm{B}\\
\mathrm{B}~ &\longrightarrow&~ \mathrm{A}\ . \label{lf146}
\end{eqnarray}
Particle B is completely inert, and does not diffuse away from its
position of creation, before it decays into A again. However note that
any point in the future is equally likely to see B change back to
A. This is very different from e.g.\ a spontaneous decay.  This
process is depicted on figure \ref{f:aab-branch}. $b$ can either come
with a positive sign, or with a negative sign. If the sign is
positive, this can be interpreted as the two particles attracting to
make the branching-process. It is clear that after some critical
threshold, the process and such the phase {\tt SAP} becomes unstable,
since the induced attraction between particles tends to make them
collapse at the same point in space and then annihilate. This leads to the runaway-flow in
phase {\tt B-1} on figure \ref{f:phases}. On the other hand, for
negative $b$, even a large $|b|$ does not lead to a collapse. This is
why on figure \ref{lambda-flow} in the case of $\lambda =\infty $ the
{\tt SAP}-phase with $a<0$ extends to $b\to -\infty $. This remains
valid for finite $\lambda $ if in the full flow-equations (\ref{lf42})
to (\ref{lf44}) $c=0$ is set from the beginning.  However the
situation for finite $\lambda $ discussed in
(\ref{lf46})--(\ref{lf47}) maps in the language of branching processes
to a finite initial ratio between $a$ and $b$, parameterized by $c =
b\rme^ {-\lambda }$, which remains uncorrected under renormalization.
The second branching-process $c$ being present,
it can render the phase {\tt SAP} unstable to {\tt B-2}. The vertex
$c$ is interpreted as 
\begin{eqnarray}
\mathrm{A}~ &\longrightarrow&~ \mathrm{C} \label{lf147}\\
\mathrm{A} + \mathrm{C}~ &\longrightarrow&~ \mathrm{A}\ . \label{lf148}
\end{eqnarray}
This means that a particle A becomes spontaneously inactive at some
time $t$. It remains at position $x$ until at some time $t'>t$ another
particle A comes by to free it. The reduced flow-equations for the
combined situation are  given in
(\ref{lf46}) and (\ref{lf47}), and lead to the instability of the
phase {\tt SAP} induced by the branching process $c$.

\section{Long-Range Elasticity}\label{LR}
Let us now study anisotropic
depinning in the case of a manifold with long range (LR) elasticity,
the elastic force in (\ref{start}) being, in Fourier:
\begin{equation}\label{lf149}
 c q^2 u_{q,t} \to (c_\alpha |q|^{\alpha} + c q^2) u_{q,t}
\end{equation}
There are now two elastic constants, the LR one $c_{\alpha }$ and the
short range (SR) one $c_{2}$, and we thus define the two dimensional
regularization-parameters,
\begin{eqnarray}
\epsilon &=& 2\alpha -d \label{lf84} \\
\kappa &=& 2- \alpha \ .\label{lf85}
\end{eqnarray}
The case of most interest corresponds to the parameters for the contact
line depinning, $d=1$, $\alpha=1$, i.e.\ $\epsilon =\kappa =1$.

\begin{figure}[t] \centerline{\Fig{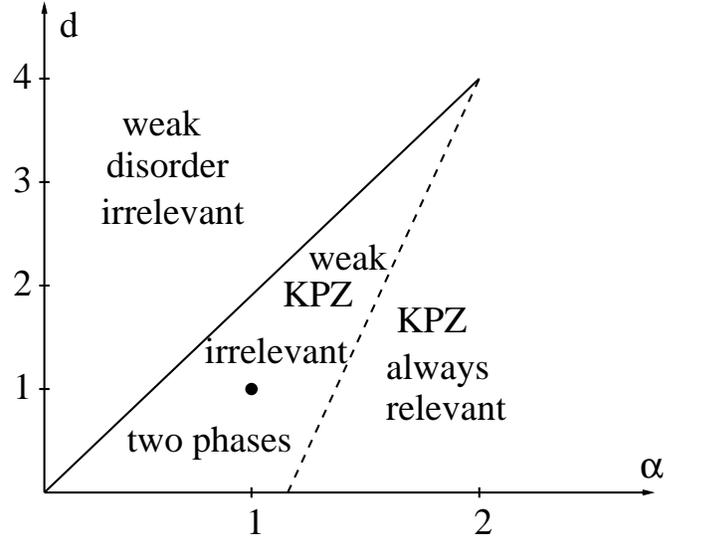}} \caption{Phase-diagram in
the ($\alpha, d$)-plane. The solid line is $\epsilon =0$. The dashed
line  corresponding to (\ref{dKPZ}) separates the domain where an
infinitesimal KPZ-term is relevant from those where it is
irrelevant. At order $\epsilon ^{2}$ this line will bend to the left,
but should not cross the point $d=1$ and $\alpha =1$.}
\label{f:KPZrel}
\end{figure}

Power counting shows that disorder is perturbatively relevant below
the critical dimension $d < d_c = 2 \alpha$. Disorder is thus relevant
for the contact line case but the crucial question we investigate here
is whether the KPZ terms are important there. Study of the contact
line depinning is usually performed within a $d=2 \alpha - \epsilon$
expansion (see Ref.~\cite{LeDoussalWieseChauve2002}) at fixed
$\alpha$. This is the solid line in figure \ref{f:KPZrel}.  However as
soon as elasticity is long range ($\kappa >0$) simple power counting
shows that the KPZ terms are perturbatively irrelevant for $d$ near
$d_c$.  Working at fixed $\alpha$ as e.g.\ $\alpha =1$ is thus not the
best method.  One alternative  is 
to study the vicinity of the point $d=4$, $\alpha=2$ and perform a
{\it double expansion} both for $\epsilon $ and $\kappa $ small.  The
idea is to determine a line $d_{\mathrm{KPZ}}(\alpha)$ in the
$(\alpha,d)$-plane below which the KPZ terms are important and must be
included.  One can determine this line near the
point $d=4$, $\alpha=2$ and, by extrapolation, find on which side of
the line lies the interesting case $\epsilon =\kappa =1$ (see figure
\ref{f:KPZrel}.).

Through the replacement $q^{2} \to q^{\alpha}$ in the propagators of the
1-loop diagrams of section \ref{RG-flow}, it is easy to derive the 1-loop
FRG equations for a general $\alpha$, in presence of a KPZ term
as in (\ref{start}).
First one obtains as usual that $c_{\alpha }$ is uncorrected to all
orders, and thus we set $c_{\alpha } = 1$ in the following. 
Defining the dimensionless couplings 
\begin{eqnarray}
 \tilde{\lambda} &=& \lambda \Lambda_\ell^{\kappa  -\zeta}\label{lf86} \\
 \tilde{\Delta}(u) &=& \Lambda_\ell^{2 \zeta - \epsilon}
\Delta(u \Lambda_\ell^{- \zeta})\label{lf87}
\end{eqnarray} 
within a Wilson scheme where $\Lambda_\ell = \Lambda \rme^{-\ell}$
is the running UV cutoff, we find the flow equations:
\begin{eqnarray} 
\partial_\ell \ln \tilde{\lambda} &=& 
\zeta - \kappa  -   \tilde{\lambda}^2 \tilde{\Delta}(0)
-  \tilde{\lambda} \tilde{\Delta}'(0^+)\label{flowlambdaLR} \\
 \partial_\ell \tilde{\Delta}(u) &=&
(\epsilon - 2 \zeta ) \tilde{\Delta}(u) 
+ u \zeta \tilde{\Delta}'(u) + 2 \tilde{\lambda}^2 \tilde{\Delta}(u)^{2} \nonumber \\
&& -\tilde{\Delta}' (u)^{2}-\tilde{\Delta}'' (u) (\tilde{\Delta}
(u)-\tilde{\Delta} (0))  \label{flowDeltaLR}\ .
\end{eqnarray}
We work to lowest order in both $\epsilon$ and $\kappa$ (and thus neglect
the small changes in the coefficients of order
$\kappa $) and define the
ratios
\begin{eqnarray} \label{lf150}
\hat \kappa = \frac{\kappa}{\epsilon} \ , \qquad \zeta_1 =
\frac{\zeta}{\epsilon}\ .
\end{eqnarray}
Of course the SR part of the elasticity is corrected:
\begin{equation} \label{3}
\partial _{\ell} \ln  \left( \frac{c_{2}}{c_{\alpha }} \right) = -\kappa
- \tilde \lambda \tilde \Delta' (0^{+}) - \tilde \lambda
^{2}\tilde \Delta (0) 
\end{equation}
and we will focus on situations where it is irrelevant (a condition 
which must be checked a posteriori).

Note that since the LR-elasticity is uncorrected, the dimensionless
variables, contrary to (\ref{lf17}), are not divided by $c_2$ but by
$c_{\alpha }=1$ and
their RG-equations thus do not contain additional contributions from
the corrections to $c_2$. As a result $\tilde{\lambda}$ has now
non-trivial corrections and the Cole-Hopf mapping no longer works, or
has to be defined with a flowing $\tilde \lambda $. 

Before embarking on a more detailed analysis let us indicate the main
behaviour we expect from Eqs.\ (\ref{flowlambdaLR}) and
(\ref{flowDeltaLR}).  For $\tilde{\lambda} = 0$ one has the usual
anisotropic depinning fixed point studied in Ref.\
\cite{LeDoussalWieseChauve2002}.  One can perform a linear stability
analysis of this FP for small $\tilde{\lambda}$. From
(\ref{flowlambdaLR}) one finds that linear stability holds provided
\begin{eqnarray}\label{lf151}
\kappa > \zeta_{\mathrm{iso}} = \frac{\epsilon}3+O (\epsilon ^{2})
\end{eqnarray}
for the non-periodic problem, and $\zeta_{\mathrm{iso}}=0$ for the
periodic case.  This is the dashed line 
\begin{equation}\label{dKPZ}
d_{\mathrm{KPZ}}(\alpha) = 5\alpha - 6
\end{equation}
represented in Fig.\ \ref{f:KPZrel}. For $d >
d_{\mathrm{KPZ}}(\alpha)$ the isotropic FP is stable.  This is the
case for the contact line depinning. On the other hand, one expects
from Eqs.\ (\ref{flowlambdaLR}) and (\ref{flowDeltaLR}) that even
then, if the value of $\tilde{\lambda}$ is large enough, the RG may
flow again to KPZ strong coupling. This is the same run-away flow as
for SR elasticity. Both fixed points should be separated by an
instable fixed point, of which we will show that it is attainable
perturbatively.  Thus for $d > d_{\mathrm{KPZ}}(\alpha)$ we expect,
and find below, {\it two phases}: one where $\tilde{\lambda}$ flows to
zero (denoted the ID phase) and one where the KPZ terms are important
(the AD phase). The question is thus to determine the basin of
attraction of each phase and the critical (i.e.\ repulsive) fixed point
which separates the two phases. Quite generally one expects a critical
value $\tilde{\lambda}^*$ below which $\tilde{\lambda}$ flows to zero
and above which it runs away.

A simple argument, confirmed by the more detailed analysis presented
below, allows to estimate $\tilde{\lambda}^*$ for small values of
$\epsilon $, i.e.\ near $d \approx d_{\mathrm{KPZ}}(\alpha)$.  Since
$\tilde{\Delta}(u)$ changes by order $\tilde{\lambda}^2$ for small
$\tilde{\lambda}$, the Eq (\ref{flowlambdaLR}) gives the critical
value:
\begin{equation}\label{lf88}
 \tilde{\lambda}^*_{ \hat \kappa}= \frac{\kappa - \zeta_{\mathrm{iso}}
}{|\tilde{\Delta}'(0^+)|}
\end{equation} 
for small $\hat \kappa - (\zeta_{\mathrm{iso}}/\epsilon)$, where
$\tilde{\Delta}'(0^+) = O(\epsilon)$ takes its (negative) value for
the isotropic depinning fixed point.

Although analysis of the full FRG-flow requires numerics, one can
obtain some analytical information on the transition between
the isotropic phase and the anisotropic strong-KPZ-coupling phase.

\subsection{Non-periodic systems}
\begin{figure}[t]
\centerline{\Fig{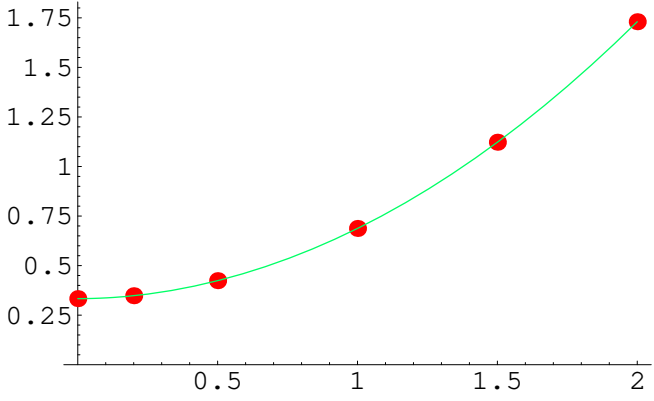}}
\caption{$\zeta_1 = \zeta/\epsilon$ as a function of $\tilde{\lambda}$ for
LR elasticity. Note that $\zeta_1
(- \tilde{\lambda}) = \zeta_1 (\tilde{\lambda} )$.}
\label{zetaoflambdaLR}
\end{figure}
\begin{figure}[t] \centerline{\Fig{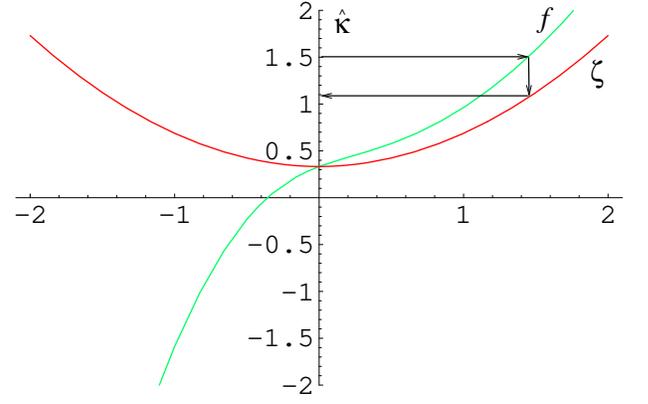}} \caption{The function $f
(x)$ (green/light) and $\zeta_1 (x)$ (red/dark) defined in the
text. One can read off $\zeta_1$ as a function of $\hat{\kappa}$ as
follows: The curve $y=f(x)$ yields $\tilde{\lambda}_{\hat \kappa}$ (x
axis) from $\hat \kappa$ (y axis) and in turn, one reads $\zeta_1(\hat
\kappa)$ (y axis) from $\tilde{\lambda}_{\hat \kappa}$ (x axis) using
the curve $y = \zeta_1 (x)$. This is indicated by the arrows. The
result is plotted in Fig. \ref{lf15a}} \label{lf13a}
\end{figure}
\begin{figure}[b]
\centerline{\Fig{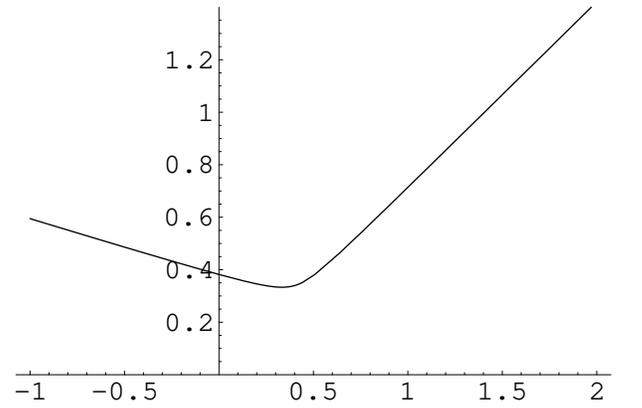}}
\caption{$\zeta_1(\hat \kappa)$, using  $\zeta (\tilde \lambda )$ and
$\tilde \lambda =f^{-1}(\hat \kappa)$} 
\label{lf15a}
\end{figure} 
Let us start with non-periodic systems and search for a perturbative
fixed point of the system (\ref{flowlambdaLR}),  (\ref{flowDeltaLR}).
Interestingly in that case, there is one, whose properties depend
continuously on $\hat \kappa = \kappa/\epsilon$.

For each value of $\hat \kappa$ we can determine the FP through the
following construction. Given the reparametrization invariance
(\ref{lf18}) of (\ref{flowDeltaLR}), we can always set 
\begin{equation}\label{haha}
\Delta
(0)=\epsilon\ ,
\end{equation}
and for each fixed value of $\tilde{\lambda}$ search
numerically for a fixed point function of Eq (\ref{flowDeltaLR}) which
decreases at infinity (short range pinning force correlations of the
random field type). Interestingly we find, through explicit numerical
integration, that there is always one such solution, denoted by
$\Delta^*_{\tilde{\lambda}} (u)$, if one tunes $\zeta$ to a value
noted $\zeta(\tilde{\lambda})$. The resulting curve
$\zeta_1(\tilde{\lambda}):= \zeta(\tilde{\lambda})/\epsilon$ is plotted
in Fig.~\ref{zetaoflambdaLR}. It starts at $\zeta(\tilde{\lambda}=0) =
1/3$ (the isotropic value) and increases as $\tilde{\lambda}$
increases.

Considering the fixed point equation of (\ref{flowDeltaLR}) at $u=0$
using (\ref{haha}) shows that the value of $\Delta' (0^{+})$ is a
simple expression:
\begin{equation}\label{lf91}
\Delta^{* \prime}_{\tilde{\lambda}} (0^{+})
=-\epsilon  \sqrt{ 1 - 2 \zeta_1(\tilde{\lambda})  + 2 \tilde{\lambda}^2 }\ .
\end{equation}
Thus, reporting this value, as well as $\Delta (0)=\epsilon$ in
Eq. (\ref{flowlambdaLR}) we see that for each value of
$\hat{\kappa}$ we can determine the value of $\tilde{\lambda}$ by
solving the  equation
\begin{equation}\label{lf90}
\hat \kappa = f(\tilde{\lambda}) \equiv \zeta_{1}(\tilde{\lambda})
- \tilde{\lambda}^2 + \tilde{\lambda} \sqrt{ 1 - 2
\zeta_1(\tilde{\lambda}) + 2 \tilde{\lambda}^2 }\ .
\end{equation}
Denoting $\tilde{\lambda}^*_{\hat \kappa}$ this solution, we
obtain the FP function $\Delta^*_{\tilde{\lambda}^*_{\hat \kappa}}
(u)$ and the value of the roughness exponent $\zeta_1(\hat \kappa)
:= \zeta_1(\tilde{\lambda}^*_{\hat \kappa})$.  Comparing
(\ref{flowlambdaLR}) and (\ref{3}) we note that the SR elastic part is
indeed irrelevant as soon as $\zeta > 0$, and thus the above analysis
is consistent.

The curves $f(\tilde{\lambda})$, $f^{-1}(\hat \kappa)$ and the
resulting $\zeta_1({\hat \kappa})=\zeta(\tilde{\lambda}^*_{\hat
\kappa})=\zeta(f^{-1}(\hat \kappa))$ are plotted in Fig.~\ref{lf13a}
 and \ref{lf15a}  respectively.

One sees that there is a solution with a positive
$\tilde{\lambda}^*_{\hat \kappa}$ only if $\hat \kappa > \kappa_c =
\frac{1}{3}$ consistent with the linear stability analysis given
above. The roughness exponent associated to this FP then increases
continuously, as shown on figure \ref{lf15a}, from $\zeta_1 =
\frac{1}{3}$ to larger values as $\hat \kappa$ increases beyond $\hat
\kappa_c$. In particular, since we are  interested in the
point $\kappa = 1, \epsilon= 1$ of the ($\alpha, d$)-plane (see figure
\ref{f:KPZrel}) it is worthwile to give the extrapolation:
\begin{equation} \label{lf93}
\zeta(\hat \kappa =1) = 0.7 \ ,
\end{equation}
and $\tilde \lambda_{\hat \kappa =1} =1.037$, values which give
the simplest extrapolation for the contact-line depinning. One should
however not expect too high a precision from this crude estimate.

Thus we have found a non-trivial FP for this problem. It continuously
depends on $\kappa/\epsilon$ and exists only for $\kappa/\epsilon >
1/3$. The simplest scenario is that this FP is associated with the
critical behaviour at the transition between the phase where KPZ is
irrelevant (isotropic depinning) and the phase where KPZ grows
(anisotropic depinning). To confirm it and check that this FP has only
one unstable direction one needs a more detailed numerical
analysis. Note that this is also indicated by an adiabatic
approximation considering (\ref{flowlambdaLR}) alone and assuming that
the disorder does not vary, which yields that the FP is repulsive if
$f'(x) >0$ and attractive if $f'(x)<0$.

\subsection{Periodic systems} In the periodic case, since $\zeta=0$ is
requested at any FP, we see that we cannot enforce the SR-elasticity
coefficient $c_2$ to scale to 0 under renormalization, since the FP
condition on $\tilde{\lambda}$ implies that (\ref{3})
vanishes. However if we start with a small ratio of $c_{2}/c_{\alpha
}$ or if the flow is such that this ratio gets small before we reach
the fixed point, then it is legitimate to neglect the effect of
$c_{2}$. We restrict our analysis to that case, and study equations
(\ref{flowlambdaLR}), (\ref{flowDeltaLR}) searching for a
FP. A more detailed numerical analysis of the flow equation is feasable.

It can  easily be seen that the form
\begin{equation}\label{lf152}
 \Delta(u) = \frac{\epsilon}{\tilde \lambda^2} (a + b e^{- u \tilde \lambda} +
c e^{u \tilde \lambda} ) 
\end{equation}
is not exactly preserved by the flow anymore (e.g. $\partial_l
\Delta(u)$ yields a term proportional to $u e^{- u \tilde \lambda}
\partial_l \tilde \lambda_l$ through variations of $\tilde \lambda_l$
which here flows).  One can still however search for exponential
fixed points since then $\tilde \lambda_l$ does not
flow. (\ref{flowDeltaLR}) yields the conditions 
\begin{eqnarray} 
 a + 2 a^2 + 4 b c &=& 0 \label{lf153}\\
 b + 4 a b + b^2 + b c &=& 0 \label{lf154} \\
 c + 4 a c + c^2 + b c &=& 0 \label{lf155}
\end{eqnarray} 
and we can set $c = b e^{-\tilde \lambda}$  to ensure periodicity
$\Delta (u)=\Delta (1-u)$. We obtain the following fixed points:
\begin{eqnarray} 
 b &=& \frac{\pm \rme^{\tilde \lambda}}{\sqrt{1 + 34 \rme^{ \tilde \lambda} +
\rme^{ 2 \tilde
\lambda}}}\label{lf156} \\ 
 a &=& - \frac{1}{4} \mp \frac{1 + \rme^{ \tilde \lambda}}{ 4 \sqrt{1 +
34 \rme^{ \tilde \lambda} + \rme^{ 2 \tilde \lambda}}}\label{lf157}
\end{eqnarray} 
as well as two others $(a=- \frac{1}{2}, b=0)$ and
$(a=0,b=0)$.  The corresponding FP condition for $\tilde{\lambda}$
gives
\begin{equation}\label{lf158}
  0 = - \hat \kappa - a - 2 b e^{ - \tilde \lambda}\ .
\end{equation}
The FP with a positive $b$ is the one of interest. It is again
presumably the boundary between the zero and strong KPZ phases. The
value of $\tilde \lambda_{\hat \kappa}$ is given by the positive root
of
\begin{equation}\label{lf159}
 4 \hat \kappa = 
1 + \frac{ \rme^{\tilde \lambda}-7}{
\sqrt{1 + 34 \rme^{ \tilde \lambda} + \rme^{2  \tilde \lambda}}}\ .
\end{equation}
which reproduces (\ref{lf88}), $\tilde \lambda_{\hat \kappa} \sim 6
\hat \kappa$, to lowest order in $\hat \kappa$. One finds that $\tilde
\lambda_{\hat \kappa}$ increases monotonically with $\hat \kappa$ and
diverges $\tilde \lambda_{\hat \kappa} \to + \infty$ as $\hat \kappa
\to \frac{1}{2}$. This suggests that for $\hat \kappa \geq
\frac{1}{2}$ only the ID phase exist.

\section{Conclusion} In this paper we have reexamined the functional
renormalization group approach to anisotropic depinning. This was
mandatory since non-analytic renormalized disorder correlators were
found to be crucial already for isotropic depinning and were neglected
in previous approaches of AD.

Indeed we have shown that the non-analyticity of  disorder arising beyond the
Larkin length is crucial to generate the KPZ-term, a first explicit
field theoretic demonstration of how these terms appear at
depinning. The resulting anomalous terms in the $\beta $-function modify
the flow compared to previous approaches in interesting ways. We found
several non-trivial fixed points and for SR elasticity a Cole Hopf
transformed theory which allows to simplify considerably perturbation
theory and indicates that the KPZ coupling $\lambda/c$ is uncorrected to
all orders.

For LR-elasticity we have found the domains of parameters belonging to
ID and AD respectively. We found that for the experimentally
interesting case of contact-line depinning, two phases exist, ID and
AD, and that the KPZ-coupling (i.e.\ the anisotropy) should be large
enough for the AD class to apply (otherwise the ID exponents is
expected
\cite{ChauveLeDoussalWiese2000a,LeDoussalWieseChauve2002a}). At the
transition a larger value of $\zeta \approx 0.7 \epsilon$ (with $\epsilon=1$
for the contact line) is obtained. This scenario could be checked in
a numerical simulation. To make
the comparison with experiments more accurate one should consider the
more involved structure for the KPZ terms unveiled in
\cite{GolestanianRaphael2001} but this can be done by methods similar
to the one introduced here.

For SR-elasticity we have found interesting new fixed points. A bit
disappointingly, they possess one unstable direction and thus
correspond to transient or critical behaviour, and not to the
asymptotic behaviour which instead is controlled by a runaway flow to
a regime not perturbatively accessible by the present method.  On the
other hand, an encouraging result is that we found a class of disorder
correlators (in the form of exponentials) which should be invariant to
all orders. These correspond to a set of branching processes which
look tantalizingly close to the ones introduced to describe reaction
diffusion and directed percolation.  More work is necessary to
understand this simpler equivalent class of theories at strong
coupling, as they may contain the key to this conjectured connection
between anisotropic depinning and directed percolation (in $d=1+1$)
and its generalizations in terms of blocking surfaces (in higher $d$)
and ultimately an understanding of the upper critical dimension for
this problem.

A posteriori, it is not suprising that the present approach yields
again a flow to strong coupling KPZ, as it does in the thermal version
of the problem \cite{Laessig1995,Wiese1998a}. It is possible that as
in the thermal problem another representation, as e.g.\ the directed
polymer, better exposes the physics and in particular what is missed
in the present approach. The corresponding formulation would be
\begin{eqnarray} \label{end1}
Z(x,t)  &=&  \int_{y(t)=x} {\cal D}[y(\tau)] \nonumber \\
&& \exp\left[ - \int_{t'}^{t} \rmd\tau  \frac{1}{4 T} \left(\frac{d y}{d
\tau} \right)^2 + \frac{1}{T} V(y(\tau) , \tau) \right] \qquad \ \label{c1}
\end{eqnarray}
i.e.\ a directed polymer in a random potential but with the choice $T
= \frac{1}{\hat \eta}$ and the additional self consistency condition:
\begin{equation}\label{end2}
V(y,\tau) = \frac{\hat \lambda}{c \hat \eta^2} F\left(y, \frac1{\hat
\lambda} \ln Z(y,\tau)\right)\ ,
\end{equation}
which relates the random potential to the pinning force and to the
free energy of the directed polymer and makes the problem analytically
far more complex. It may possess similar physics and thus be
amenable to some extended FRG approach which would better account (as
it does for the thermal problem) for the coarse grained correlations
in the $y$ direction a property clearly not taken into account by the
present method, which treats correctly only correlations in the $\ln
Z$ space.

\begin{acknowledgments}
It is a pleasure to thank M.~Kardar, W.~Krauth, A.~Ludwig and A.~Rosso for
stimulating discussions.
\end{acknowledgments}

\appendix

\section{Diagrams} \label{app-corrections} We use the following model
setting $c=\eta=1$ to simplify notations
\begin{eqnarray}\label{lf17a}
S&=&\int_{x,t} \eta \hat{u}\dot u -c \hat{u}\Delta u -\lambda
\hat{u} (\nabla u)^{2} \\ 
&& -\frac12
\int_{x,t,t'}\hat{u}_{xt}\hat{u}_{xt'}\Delta(u_{xt}-u_{xt'})   \label{lf94}
\end{eqnarray}
One also has to specify a cut-off procedure. For convenience, we chose
to put a mass-term. This is justified at 1-loop order since the
results are universal, i.e.\ cut-off independent. At second order, one
would have to be more careful and use, e.g. an external momentum IR cutoff.

Many of the diagrams which we need are identical to the driven
manifold problem at $\lambda =0$. These diagrams are detailed in
\cite{LeDoussalWieseChauve2002}.  The {\em new} diagrams are
\begin{eqnarray}
\diagram{KPZ1}&=& 2 \int_{0<t<t'}\int_{k} \rme^{-t (k^{2}+m^{2})} 
\rme^{-t' (k^{2}+m^{2})} |t-t'| k^{2} \nn \\
&&\hphantom{ 2 \int_{0<t<t'}\int_{k}} \times \Delta' (0^{+})\hat{u}
\dot{u}\nn\\  
&=& 2\lambda \Delta' (0^{+})\hat{u}\dot u\int_{k}\int_{t'}\rme^{-t'
(k^{2}+m^{2})} \frac{k^{2}}{(k^{2}+m^{2})^{2}}\nn \\ 
&& \qquad \qquad \times \left(\rme^{-t' (k^{2}+m^{2})}-1-
(k^{2}+m^{2})t' \right)\nn\\ 
&=& 2\lambda \Delta' (0^{+})\hat{u}\dot u\nn \\
&&\times \int_{k }\frac{k^{2}}{2
(k^{2}+m^{2})^{3}}-\frac{k^{2}}{(k^{2}+m^{2})^{3}}+
\frac{k^{2}}{(k^{2}+m^{2})^{3}}\nn\\  
&=&\lambda \Delta' (0^{+})\hat{u}\dot u\int_{k}
\frac{k^{2}}{(k^{2}+m^{2})^{3}} \nonumber \\ 
&=& \lambda \Delta' (0^{+})\hat{u}\dot u \int_{k} \frac{k^{2}}{(
k^{2}+m^{2})^{3}} \label{lf95}
\end{eqnarray}
\begin{equation} \label{lf96}
\diagram{KPZ2}=-\frac{4}{d} \Delta' (0^{+}) \lambda ^{2} \int
\frac{k^{2}}{(k^{2}+m^{2})^{3}}  \ \hat{u} (\nabla u)^{2}
\end{equation}
\begin{eqnarray}
\diagram{KPZ3}&=& 2 \Delta' (0^{+}) \lambda
 \int_{k}\frac{( k_{\alpha }+p_{\alpha })p_{\alpha }}{(
(k+p)^{2}+m^{2}) (k^{2}+m^{2})}\nonumber \\ 
&\approx &  2 \Delta' (0^{+}) \lambda 
\int_{k} \frac{p^{2}}{(k^{2}+m^{2})^{2}}-\frac{2
(kp)^{2}}{(k^{2}+m^{2})^{3}} \nonumber \\ 
&=& \frac{2 (2-d)}{d} \Delta' (0^{+}) \lambda
\int_{k}\frac{1}{(k^{2}+m^{2})^{2}}  \hat{u}\Delta{u} 
\nn \\ \label{lf97}
\end{eqnarray}
(Note that $\Delta \leftrightarrow  - p^{2}$.) Dots indicate omitted
subleading terms. 
\begin{equation}
\diagram{KPZ4}=-\frac{8}{d}\Delta (0)\lambda
^{3}\int_{k}\left[\frac{k^{4}}{(k^{2}+m^{2})^{4}}+\dots  \right]
\hat{u} (\nabla u)^{2} \label{lf98} 
\end{equation}
\begin{eqnarray}
\diagram{KPZ5} 
&=& -4 \Delta (0) \lambda ^{2}\int_{k} \frac{k
(k+p)\,(kp)}{(k^{2}+m^{2})^{2} ((k+p)^{2}+m^{2})}\nonumber \\ 
&=& - 4 \Delta (0)\lambda
^{2}\int_{k}\frac{(kp)^{2}}{(k^{2}+m^{2})^{3}}-2 \frac{k^{2}
(kp)^{2}}{(k^{2}+m^{2})^{4}}\nonumber \\ 
&=&-\frac{4}{d}\Delta (0) \lambda ^{2} \left(\int_{k}
\frac{1}{(k^{2}+m^{2})^{2}}+\dots  \right)\left(\hat{u}\Delta u \right)
\nonumber \\ \label{lf99}
\end{eqnarray}
\begin{equation} \label{lf100}
\diagram{KPZ6}=\frac{4}{d}\Delta (0) \lambda ^{3} \int_{k}
\frac{k^{4}}{(k^{2}+m^{2})^{4}} \ \hat{u} (\nabla u)^{2}
\end{equation}
\begin{equation} \label{lf101}
\diagram{KPZ7}= \Delta (u_{xt}-u_{xt'})^{2}\lambda ^{2}
\int_{k} \frac{(k^{2})^{2}}{(k^{2}+m^{2})^{4}} \hat{u}_{xt}\hat{u}_{xt'}
\end{equation}
\begin{equation} \label{lf102}
\diagram{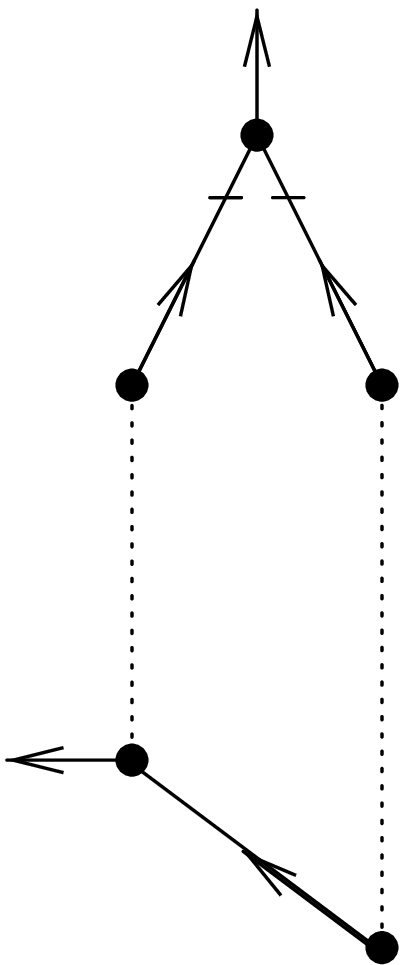}+\diagram{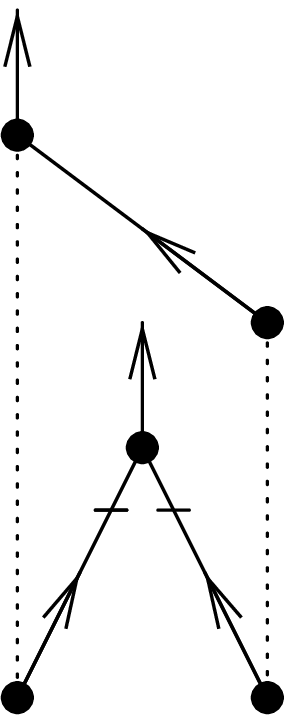}=0
\end{equation}
Therefore, we have the following corrections to $c$, $\eta$, $\lambda
$ and $\delta $ (setting $I:=\int_{k}\frac{1}{(k^{2}+m^{2})^{2}}$,
dropping finite terms in $\epsilon $, but for the moment keeping the
explicit $d$-dependence)
\begin{eqnarray}
\delta c /c &=& \left[\frac{2 (2-d)}{d}\Delta' (0^{+})\lambda
-\frac{4}{d}\Delta (0)\lambda ^{2} \right] I \label{lf115}\\ 
\delta \eta/\eta&=& -\lambda \Delta'
(0^{+}) I \label{lf103} \\
\delta \lambda /\lambda &=&\left[- \frac{4}{d}\lambda \Delta'
(0^{+})-\frac{8}{d}\Delta 
(0)\lambda ^{2}+\frac{4}{d}\Delta (0)\lambda ^{2} \right]  I\nonumber \\
\label{lf116}\\ 
\delta \Delta (u) &=& 2 \Delta (u)^{2} \lambda ^{2} I \label{lf104}
\end{eqnarray}
The coupling-constant is $\hat{\lambda }:=\lambda /c$. Note that its
flow vanishes at leading order in $1/\E $.  We now check cancellations
beyond the leading order. We use
\begin{equation} \label{lf105}
\int_{k}=  A_{d}\int_{0}^{\infty}
\rmd k k^{d-1}
\end{equation}
The two diagrams proportional to $\Delta' (0^{+})$ are:
\begin{eqnarray}
\diagram{KPZ2} &\longrightarrow& -\int_{k}
\frac{4}{d}\frac{k^{2}}{(k^{2}+m^{2})^{3}}\nonumber \\
&=& A_{d} \frac{\left( -2 + d \right) \,\pi \,\csc (\frac{d\,\pi
}{2})}
  {4\, m^{\epsilon }} \label{lf117}\\
\diagram{KPZ3}& \longrightarrow& \int_{k}\frac{-4\,k^2}{d\,{\left( k^2
+ m^2 \right) }^3} +  
  \frac{2}{{\left( k^2 + m^2 \right) }^2} \nonumber \\
&=& A_{d}\frac{\left( 2 - d \right) \,\pi \,
      \csc (\frac{d\,\pi }{2})  }{4\, m^\epsilon } \label{lf106}
\end{eqnarray}
The sum (which gives the renormalization of $\hat{\lambda }$) exactly
vanishes.

The corrections proportional to $\Delta (0)$ are 
\begin{eqnarray}
\diagram{KPZ4} &\longrightarrow& \int_{k}\frac{-8\,k^4}{d\,{\left( k^2
+ m^2 \right) }^4}\nonumber \\ 
 &=& A_{d}\frac{\left( -4 + d^2 \right) \,\pi \,
    \csc (\frac{d\,\pi }{2})}{12\,
    m^{\epsilon }} \label{lf107}
\\
\diagram{KPZ6}& \longrightarrow& \int_{k}\frac{4\,k^4}{d\,{\left( k^2
+ m^2 \right) }^4}\nonumber \\ 
 &=&A_{d} \frac{\left( 4 - d^2 \right) \,\pi \,
    \csc (\frac{d\,\pi }{2})}{24\, m^{\epsilon }} \label{lf108}
\\
\diagram{KPZ5} &\longrightarrow& \int_{k}\frac{8\,k^4}{d\,{\left( k^2
+ m^2 \right) }^4} -  
  \frac{4\,k^2}{d\,{\left( k^2 + m^2 \right) }^3}\nonumber \\
 &=&A_{d}
\frac{ \left( d - 2 \right) \,
      \left( d-1 \right) \,\pi \,\csc (\frac{d\,\pi }{2})
       }{12\,m^{\epsilon }}\nonumber \\\label{lf118}
\end{eqnarray}
The sum of the above three terms is
\begin{equation}
A_d \frac{ \left( 4 - d \right) \,
      \left( d-2 \right) \,\pi \,\csc (\frac{d\,\pi }{2})
       }{24\,m^{\epsilon }} \label{lf109}
\end{equation}
Note that
\begin{equation}
\pi \,\csc (\frac{d\,\pi }{2}) = \frac{2}{d-4} + \frac{{\pi
}^2\,\left(  d-4 \right) } 
   {12} + \frac{7\,{\pi }^4\,{\left( d-4 \right) }^3}
   {2880} + \dots  \label{lf110}
\end{equation}
So, working in a massive scheme, there are corrections at order $\E$,
compared to the leading term which would be $1/\E$. We see that the
fixed point of Stepanow \cite{Stepanow1995} is --even if one would
accept his scheme-- incorrect.  However, as we have already stated
above, one should do the calculations in a massless scheme.

\section{Long range disorder}

\label{lrdis}

In this Appendix we give a quick study of the case with
long range disorder in internal space $x$. We show that one
recovers the Flory estimate of Section IV in the case of
isotropic depinning. For anisotropic depinning we find a 
runaway flow and cannot conclude. 

We study
\begin{eqnarray}
S_{\mathrm{DO}}&=& \frac{1}{2} \int_{xtx't'} \hat Z_{x t} \hat Z_{x'
t'} \hat \lambda^2 \Delta(u_{xt} - u_{x't'}) f(x-x')\qquad
\label{lf119} \\ 
f(x) &\sim& x^{- \alpha} \label{lf111}
\end{eqnarray}
We find the FRG equation for the LR disorder:
\begin{equation}\label{lf112}
 \partial \Delta = \epsilon \Delta + \Delta(0) \Delta'' + (2 - \mu)
(\lambda \Delta'(0^+) + \lambda^2 \Delta(0) ) \Delta 
\end{equation}
with $\epsilon=4 - \alpha$, $d$ large enough ($d >\alpha$ or more). We
have absorbed $\epsilon A$ in $\Delta$ with:
\begin{equation}\label{lf113}
 A = \int_q C(q)^2 f(q)
\end{equation}
This is because the graphs leading to two $\Delta(u)^2$ functions or
more do not contribute. This remains true to all orders, inspection
for $\lambda=0$ shows that to two or three loops no corrections arise,
except anomalous terms (which, as we will see are not needed as we
find analytic fixed points). So for $\lambda=0$ the one loop result is
probably exact to all orders.

The coefficient $\mu$ comes from the corrections to the gradient term:
\begin{eqnarray}
&&  (\lambda \Delta'(0^+) + \lambda^2 \Delta(0) ) B\label{lf120} \\
&& B = \frac{1}{2 d} \int_x x^2 f(x) C(x)\label{lf121} \\
&& \mu = 2 B/A = \frac{2 (d-4)}{d}\label{lf122}
\end{eqnarray}
with $\alpha=4$, $d>4$  (note that it goes to 2 when
$d$ goes to infinity).  

One easily finds the fixed points for $\lambda=0$. For periodic
disorder one has:
\begin{eqnarray}
\Delta(u) &=& g \cos(2 \pi u)\label{lf125} \\
\partial g &=& \epsilon g - (2 \pi)^2 g^2\label{lf126}
\end{eqnarray}
The correlations are:
\begin{equation}\label{lf124}
\langle  u u\rangle  = \hat \Delta(0) \frac{f(q)}{q^4} \sim q^{-(d + 2 \zeta)}
\end{equation}
with $\zeta=\epsilon/2$ as if $\Delta(0)$ was uncorrected.

For non-periodic disorder, rescaling $\Delta$ gives:
\begin{eqnarray}
  \tilde{\Delta}(u) &=&  \tilde{\Delta}(0) \rme^{- \epsilon u^2/(6
\tilde{\Delta}(0))} \label{lf127}\\ 
 \zeta&=& \epsilon/3\label{lf128}
\end{eqnarray}
Thus 
the Flory estimate is exact. Note that the fact that the LR correlator
$\tilde{\Delta}(u)$ is analytic is not puzzling, since it generates in
turn a SR part which should be nonanalytic in order to e.g. successfully
generate a depinning threshold force.

On the other hand for $\lambda>0$ we find that:
\begin{eqnarray}
 \Delta(u) &=& g \cos(2 \pi u)\label{lf129} \\
 \partial g &=& \epsilon g - (2 \pi)^2 g^2 + \gamma \lambda^2 g^2\label{lf130}
\end{eqnarray}
and there is thus a critical $\lambda$ beyond which there is no  fixed
point. This seems also to be the case  for  RF. Because of this
runaway flow we cannot conclude.




\begin{thebibliography}{10}

\bibitem{Kardar1997}
M.~Kardar,
\newblock {\em Nonequilibrium dynamics of interfaces and lines},
\newblock Phys. Rep. {\bf 301} (1998)   85--112.

\bibitem{LeDoussalGiamarchi1997}
P.~Le Doussal and T.~Giamarchi,
\newblock {\em Moving glass theory of driven lattices with disorder},
\newblock Phys. Rev. {\bf B 57} (1998)   11356--11403.

\bibitem{AmaralBarabasiStanley1994}
L.A.N. Amaral, A.~L. Barabasi  and H.E. Stanley,
\newblock {\em Critical dynamics of contact line depinning},
\newblock Phys. Rev. Lett. {\bf 73} (1994).

\bibitem{TangKardarDhar1995}
L.-H. Tang, M.~Kardar  and D.~Dhar,
\newblock {\em Driven depinning in anisotropic media},
\newblock Phys. Rev. Lett. {\bf 74} (1995)   920--3.

\bibitem{AlbertBarabasiCarleDougherty1998}
R.~Albert, A.-L. Barabasi, N.~Carle  and A.~Dougherty,
\newblock {\em Driven interfaces in disordered media: determination of
  universality classes from experimental data},
\newblock Phys. Rev. Lett. {\bf 81} (1998)   2926--9.

\bibitem{TangLeschhorn1992}
L.-H. Tang and H.~Leschhorn,
\newblock {\em Pinning by directed percolation},
\newblock Phys. Rev. A {\bf 45} (1992)   R8309--12.

\bibitem{BuldyrevBarabasiCasertaHavlinStanleyVicsek1992}
S.V. Buldyrev, A.-L. Barabasi, F.~Caserta, S.~Havlin, H.E. Stanley  and
  T.~Vicsek,
\newblock {\em Anomalous interface roughening in porous media: experiment and
  model},
\newblock Phys. Rev. A {\bf 45} (1992)   R8313--16.

\bibitem{GlotzerGyureSciortinoConiglioStanley1994}
S.C. Glotzer, M.F. Gyure, F.~Sciortino, A.~Coniglio  and H.E. Stanley,
\newblock {\em Pinning in phase-separating systems},
\newblock Phys. Rev. E {\bf 49} (1994)   247--58.

\bibitem{StanleyBuldyrevGoldbergerHaussdorfMietusPengSciortinoSimons1992}
H.E. Stanley, S.V. Buldyrev, A.L. Goldberger, A.L. Haussdorf, J.~Mietus, C.-K.
  Peng, F.~Sciortino  and M.~Simons,
\newblock {\em Fractal landscapes in biological systems: long-range
  correlations in dna and interbeat heart intervals},
\newblock Physica {\bf A 191} (1992).

\bibitem{BuldyrevHavlinStanley1994}
S.V. Buldyrev, S.~Havlin  and H.E. Stanley,
\newblock {\em Anisotropic percolation and the $d$-dimensional surface
  roughening problem},
\newblock Physica {\bf A 200} (1993).

\bibitem{HavlinAmaralBuldyrevHarringtonStanley1995}
S.~Havlin, L.A.N. Amaral, S.V. Buldyrev, S.T. Harrington  and H.E. Stanley,
\newblock {\em Dynamics of surface roughening with quenched disorder},
\newblock Phys. Rev. Lett. {\bf 74} (1995)   4205--8.

\bibitem{BarabasiGrinsteinMunoz1996}
A.-L. Barabasi, G.~Grinstein  and M.A. Munoz,
\newblock {\em Directed surfaces in disordered media},
\newblock Phys. Rev. Lett. {\bf 76} (1996)   1481--4.

\bibitem{DSFisher1986}
D.S. Fisher,
\newblock {\em Interface fluctuations in disordered systems: {$5-\epsilon$}
  expansion},
\newblock Phys. Rev. Lett. {\bf 56} (1986)   1964--97.

\bibitem{NattermanStepanowTangLeschhorn1992}
T.~Nattermann, S.~Stepanow, L.H. Tang  and H.~Leschhorn,
\newblock {\em Dynamics of interface depinning in a disordered medium},
\newblock J. Phys. II France {\bf 2} (1992)   1483--1488.

\bibitem{LeschhornNattermannStepanow1996}
H.~Leschhorn, T.~Nattermann, S.~Stepanow  and L.H. Tang,
\newblock {\em Driven interface depinning in a disordered medium},
\newblock Ann. Physik {\bf 6} (1997)   1--34.

\bibitem{NarayanDSFisher1992a}
O.~Narayan and D.S. Fisher,
\newblock {\em Dynamics of sliding charge-density waves in 4- epsilon
  dimensions},
\newblock Phys. Rev. Lett. {\bf 68} (1992)   3615--18.

\bibitem{NarayanDSFisher1993a}
O.~Narayan and D.S. Fisher,
\newblock {\em Threshold critical dynamics of driven interfaces in random
  media},
\newblock Phys. Rev. B {\bf 48} (1993)   7030--42.

\bibitem{ChauveLeDoussalWiese2000a}
P.~Chauve, P.~Le Doussal  and K.J. Wiese,
\newblock {\em Renormalization of pinned elastic systems: How does it work
  beyond one loop ?},
\newblock Phys. Rev. Lett. {\bf 86} (2001)   1785--1788.

\bibitem{LeDoussalWieseChauve2002}
P.~Le Doussal, K.J. Wiese  and P.~Chauve,
\newblock {\em 2-loop functional renormalization group analysis of the
  depinning transition},
\newblock cond-mat\slash {\bf 0205108} (2002).

\bibitem{RossoKrauth2001b}
A.~Rosso and W.~Krauth,
\newblock {\em Origin of the roughness exponent in elastic strings at the
  depinning threshold},
\newblock Phys. Rev. Lett. {\bf 87} (2001)   187002.

\bibitem{RossoKrauth2002}
A.~Rosso and W.~Krauth,
\newblock {\em Roughness at the depinning threshold for a long-range elastic
  string},
\newblock Phys. Rev. E {\bf 65} (2002)   025101/1--4.

\bibitem{Stepanow1995}
S.~Stepanow,
\newblock {\em Dynamics of growing interfaces in a disordered medium: the
  effect of lateral growth},
\newblock J. Phys. II (France) {\bf 5} (1995)   11--17.

\bibitem{Laessig1995}
M.~L\"assig,
\newblock {\em On the renormalization of the {Kardar-Parisi-Zhang} equation},
\newblock Nucl. Phys. {\bf B 448} (1995)   559--574.

\bibitem{Wiese1998a}
K.J. Wiese,
\newblock {\em On the perturbation expansion of the {KPZ}-equation},
\newblock J. Stat. Phys. {\bf 93} (1998)   143--154.

\bibitem{GolestanianRaphael2001}
R.~Golestanian and E.~Raphael,
\newblock {\em Relaxation of a moving contact line and the Landau-Levich
  effect},
\newblock Europhys. Lett. {\bf 55} (2001)   228--34.

\bibitem{LeDoussalWieseChauve2002a}
P.~Le Doussal, K.~Wiese  and P.~Chauve,
\newblock {\em Two loop {FRG} study of pinned manifolds},
\newblock in preparation.

\bibitem{NattermannRenz1989}
T.~Nattermann and W.~Renz,
\newblock {\em Diffusion in a random catalytic environment, polymers in random
  media, and stochastically growing interfaces},
\newblock Phys. Rev. {\bf A 40} (1989)   4675.

\bibitem{LeDoussalWiese2001}
P.~Le Doussal and K.J. Wiese,
\newblock {\em Functional renormalization group at large {$N$} for random
  manifolds},
\newblock cond-mat\slash {\bf 0109204} (2001).

\end{thebibliography}

\end{document}